\documentclass[preprint,floatfix,aip,rsi,amsfonts,amsmath,amssymb,superscriptaddress]{revtex4-1}%

\usepackage[english]{babel}
\usepackage[T1]{fontenc}
\usepackage[utf8x]{inputenc}
\usepackage{libertine}

\usepackage{epsfig}
\usepackage{graphicx} 
\usepackage{stackengine}
\usepackage{subcaption}
\usepackage{color}
\usepackage{xcolor}

\usepackage{amsmath,amssymb}
\usepackage{array}
\usepackage{units}
\usepackage{siunitx}

\usepackage{url}

\begin{document}

\title{Optimized Setup for 2D Convection Experiments in Thin Liquid Films}
\author{Michael Winkler}
\affiliation{ University of Potsdam, Institute of Physics and Astronomy, 14476 Potsdam, Germany }

\author{Markus Abel}
\affiliation{ Ambrosys GmbH, 14473 Potsdam, Germany}
\affiliation{ University of Potsdam, Institute of Physics and Astronomy, 14476 Potsdam, Germany }

\date{\today}

\begin{abstract}
 We present a novel experimental setup to investigate two-dimensional thermal convection in a freestanding thin liquid film. We develop a setup for the reproducible generation of freestanding thin liquid films. Such films can be produced in a controlled way on the scale of 5 to 1000 nanometers. Our primary goal is to investigate the statistics of reversals in Rayleigh-B\'enard convection with varying aspect ratio; here numerical works are quite expensive and 3D experiments prohibitively complicated and costly. However, as well questions regarding the physics of liquid films under controlled conditions can be investigated, like surface forces, or stability under varying thermodynamical parameters. The thin liquid film has a well-defined and -chosen chemistry in order to fit our particular requirements, it has a thickness to area ratio of approximately \num{e8} and is supported by a frame which is adjustable in height and width to vary the aspect ratio $\Gamma=0.16$ to $10$. The top and bottom frame elements can be set to specific temperature within T= \SIrange{15}{55}{\celsius}.
The ambient parameters of the thin film are carefully controlled to achieve reproducible results and allow a comparison to experimental and numerical data.

\end{abstract}

\pacs{47.52.+j,47.55.P-,68.15.+e,47.27.wj,47.55.pb,47.27.-i,82.70.Rr,47.51.+a}

\maketitle

\section{\label{intro} Introduction}
Convection is of primordial importance with many applications from astrophysics to nanophysics, i.e. on the large scale \cite{ahlers2009heat} and on the small scale \cite{LohseXia2010RBC}.
Here, we present a two-dimensional analogue to the widely studied three-dimensional Rayleigh-Bénard convection cells: A vertically free-standing thin liquid film spanning across a metal frame is heated from below and cooled from above. The advantage of a two-dimensional setup is that the geometry can be varied with relative ease and the cost of construction and operation is far less compared to three-dimensional setups. However, these advantages are traded in for more complicated boundary conditions.

Our focus lies on the investigation of Rayleigh-Bénard convection, one paradigmatic system in hydrodynamics. It consists of fluid confined between a heat source with gravity acting parallel to the temperature gradient. It is widely used to study pattern formation and transitions between laminar and turbulent flow. In the last years, the question of reversals in Rayleigh-B\'enard convection has been studied numerically and experimentally \cite{Sugiyama-Ni-Stevens-Chan-Zhou-Xi-Sun-Grossmann-Xia-Lohse-10,mishra2011dynamics,PhysRevLett.102.144503,araujo2005wind}.
The setup is designed flexible in order to vary geometry and inflow boundary conditions, in particular the change of the flow characteristics with aspect ratio $\Gamma$ (horizontal versus vertical dimension) and with temperature difference. Thereby, we cover fundamental and applied aspects, extending
thin film experimental techniques to a very important non-equilibrium situation.

First steps towards experiments with stationary films and thermal driving have been reported, but the chemical composition and evaporation from the film have neither been controlled nor monitored
\cite{martin1998spectra,zhang2005density,seychelles2010intermittent}. One reason might be the relatively naive choice of chemicals (commercial detergents) which contain an ensemble of unknown surfactants and may contain anti-foaming agents which limit film stability. In contrast, many experiments with extremely well controlled chemistry and environment have been done on static thin films with a focus on the detailed understanding of interfacial physics, including surfactants and the chemical details of the liquid, \cite{exerowa1998foam,ivanov1988thin,PhysRevE.76.056306,stubenrauch2004stability}.

The presented measurement setup was designed with a precise control of the solution composition and ambient parameters in mind. As evaporation from the thin film, especially in heat driven convection scenarios can dramatically change the surface properties of a thin liquid film, a well-controlled experiment is needed.

It is possible to stably observe convection patterns for film thicknesses of  \SIrange{50}{500}{\nano\metre}, over a long time. Consequently, the developed apparatus serves as a
two-dimensional, well-controlled experiment using
\textit{stable, free-standing} thin liquid films to study two-dimensional
Rayleigh-B\'enard convection. Typical Rayleigh numbers of \numrange{e6}{e7} are
reached with a centimeter size experiment.

Open questions concern the change of fluid properties with thickness, and the detailed influence of additional chemical forces (e.g. from surfactants), in comparison to classical fluid dynamics.

\subsection{Thin Liquid Films}
Thin film dynamics is governed by \textit{gravitational}, \textit{capillary} and \textit{interfacial} forces. The drainage of thin films induced by the capillary pressure $P_C$ is countered by interactions between the film surfaces. These interactions are combined in the disjoining pressure $\Pi$ which acts acts perpendicular to the interfaces, thus balancing $P_C$. A quasistatic equilibrium is established when $P_C = \Pi$. The relevant interactions constituting $\Pi$ are electrostatic, van der Waals (vdW), and steric forces \cite{bergeron1999forces,derjaguin1989theory} and strongly depend on the distance between the interacting surfaces.

Whereas films on substrates are established in industry and research \cite{reiter1998artistic}, freestanding thin liquid films (hereafter TLFs) still provide a challenge in experiments and theory alike.
Consequently, the study of TLFs is central to many current scientific activities, e.g. \cite{kellay2011turbulence,yunker2011suppression,davey2010enantiomer,Prudhomme-Khan-96,exerowa1998foam,vermant2011fluid}.
We contribute by presenting a novel experimental setup to analyze convection in \textit{vertically oriented, freestanding, thermally forced, non-equilibrium} TLFs.

As a result of the force balance, two stable equilibria may occur, depending on the chemical composition of bulk solution and chosen surface active agents (surfactants): Common Black Films (CBF) with a thickness of more than \SI{10}{\nano\metre} are formed when electrostatic interactions balance the dominant van der Waals force, and, of course, gravity and capillarity \cite{derjaguin1989theory,Verwey-Overbeek-48}; Newton Black Films (NBF) are stable  with  a thickness of less than \SI{10}{\nano\metre}, due to repulsive short range steric forces \cite{jones1966stability,israelachvili1991intermolecular}.
Depending on the chemical composition of the liquid the film will reach a NBF or CBF \cite{exerowa1998foam,ivanov1988thin}. In order to distinguish these cases it is essential to have well controlled experimental conditions: filtered, deionized water, pure surfactants and control over the ambient parameters (temperature and humidity).

For the non-equilibrium dynamical behaviour, studies have so far focused on the role of gravitational or capillary forces in undisturbed, freestanding TLFs \cite{ivanov1988thin,Ivanov-80}. The effect of additional forces has been studied by a few authors, mostly for micrometer-thick systems \cite{zhang2005velocity,seychelles2008thermal}. The formation of TLFs, aided by thermal forcing is an open issue on the nanoscale. For the most part of the presented convection experiments the TLF will be in a quasi-steady but transient state. The convection prevents stratification, therefore the TLF does not reach its equilibrium thickness (CBF or NBF): the slow, gravity driven thinning is counteracted and a more uniform thickness is maintained. In this quasi-steady state convection is vigorous and can last indefinitely without the formation of an equilibrium phase layer.
Additionally, the convection can be triggered after the TLF has reached its equilibrium NBF state so that forced convection in a sub \SI{50}{\nm} TLF can be observed. To our knowledge our setup is the first to make this observation possible. As the CBF or NBF show almost no reflection of (visible) light due to the particular interference conditions (the film can be much thinner than visible light), data acquisition and analysis is challenging.

The driving mechanism in this setup is a temperature gradient antiparallel to gravity which results in a buoyancy-driven convection, analogue to classic Rayleigh-Bénard convection.
Please note that no Marangoni type flows are present in this setup ($Ma\simeq 0$).
The TLF is bound by the frame, such that no conventional Marangoni effect
occurs, since there is no free interface and consequently no thermocapillary
effect occurs. However, there is an interface with the holding frame, subject to
no-slip boundary conditions, which hinders any tangential flow.
More heuristically,  a Marangoni effect would drive a flow orthogonal to the vertically oriented
surface. We, however observe a flow in the plane of the film, i.e., transversal
to the surface.
Additionally, the surface is mobile and therefore no surfactant concentration gradients arise which could lead to a Marangoni flow \cite{breward2002drainage,howell2005absence}. No surface stresses can arise and the surface tension is constant\cite{muruganathan2004foam} at $\sigma= 34 \cdot 10^{-3} \,\unitfrac{N}{m}$.

\subsection{Rayleigh-Bénard-Convection in Thin Liquid Films}
One of the long-standing, very important questions in convection problems concerns the understanding
of the convective-to-conductive heat transport, measured by the Nusselt number, $Nu = Nu(Ra,Pr,\Gamma)$,  and of the momentum transport, characterized by the Reynolds number, $Re=Re(Ra,Pr,\Gamma)$, where $\Gamma$ denotes the aspect ratio of the container. The scaling of Nu(Ra), Re(Ra) has been studied, mainly in water and air, few for liquid metals. The corresponding range of Prandtl numbers is relatively limited. Experimental studies on the aspect ratio,  $\Gamma$, exist, are however cost intensive, e.g. for the ''barrel of Ilmenau'' \cite{thess2001barrel},
where a variation of $0<\Gamma\lesssim 1$ is possible by controlling the cool top of a cylindrical vessel
\cite{PhysRevE.75.016302,resagk2006oscillations}.

Under certain conditions, a flow develops a large-scale circulation (LSC). It is observed experimentally
and numerically in 3D and 2D  that this LSC breaks down and/or reverts in a regular or irregular way, depending on $Ra$ and geometry \cite{PhysRevLett.95.084502,bureau2010nonlinear,
Sugiyama-Ni-Stevens-Chan-Zhou-Xi-Sun-Grossmann-Xia-Lohse-10,stevens2009transitions,PhysRevE.83.036307,CambridgeJournals:7996276}.
Recent 2D numerical studies of Lohse et al. indicate that the dynamics of reversals strongly depends on $\Gamma$ \cite{Sugiyama-Ni-Stevens-Chan-Zhou-Xi-Sun-Grossmann-Xia-Lohse-10}.

In contrast to 3D convection experiments, there is relatively little done {\em experimentally} on 2D, or quasi-2D systems. Initial convection studies of foam films were carried out by Mysels \cite{mysels1959soap}, later by Couder \cite{couder1989hydrodynamics}. A theoretical framework was established by Chomaz \cite{chomaz2001dynamics} and Bruinsma \cite{bruinsma1995theory}.  Typically, free-standing TLFs are employed as fluid systems, where a collapse of the film is prevented by repumping liquid in order to compensate for losses due to evaporation, and of course gravitational flow downwards \cite{kellay1995experiments}. This results in relatively uncontrolled boundary conditions, since the free surface can vary in thickness and can even develop instabilities (ripples, etc.).  However, the system is well suited for experiments on 2D turbulence, because for Reynolds numbers up to at least \num{e8} the thickness
is well below the Kolmogorov length, which marks the scale above which no turbulence occurs, and dissipation dominates the momentum transfer \cite{Frisch-95} . In 2D, the momentum is transferred from small to large scales by the so-called "inverse cascade" in contrast to 3D, where the opposite is observed as the "direct cascade" \cite{Frisch-95,kellay2004dispersion,jullien1999richardson}. In extremely large TLFs  (with heights of several meters) information about the inverse cascade is observed experimentally by particle imaging and laser doppler velocimetry. A well explored topic is the evolution of grid turbulence with seeded sub-micron particles \cite{martin1998spectra,vorobieff1999soap,rivera1998turbulence,  shakeel2007decaying,schnipper2009vortex}.

For a two--dimensional TLF, one can assume an effective decoupling of buoyancy-caused advection and surface-related relaxation due to disjoining and capillary pressure. Of course, in a complete, 3D description basically all forces are coupled.  For very small length scales, a molecular description is necessary, however the hydrodynamic approximation holds well down to a few molecular layers. These issues and further details can be read in \cite{ChemSocRev39-2010}, where, however, mainly fluid confined by solid walls are considered. More fundamental discussion is given in \cite{Physics.3.73,bocquet_nanofluidics_2010,PhysRevLett.105.106101,bureau2010nonlinear}.
It is expected that the scales considered here (\SI{>5}{\nano\metre}) are sufficiently described by hydrodynamic equations. This assumption is not completely unreasonable as $Ra \sim \num{e7}$ and thus other forces are smaller by orders of magnitude.

\section{System Description}
\begin{figure}
  \centering
  \includegraphics[width=\textwidth]{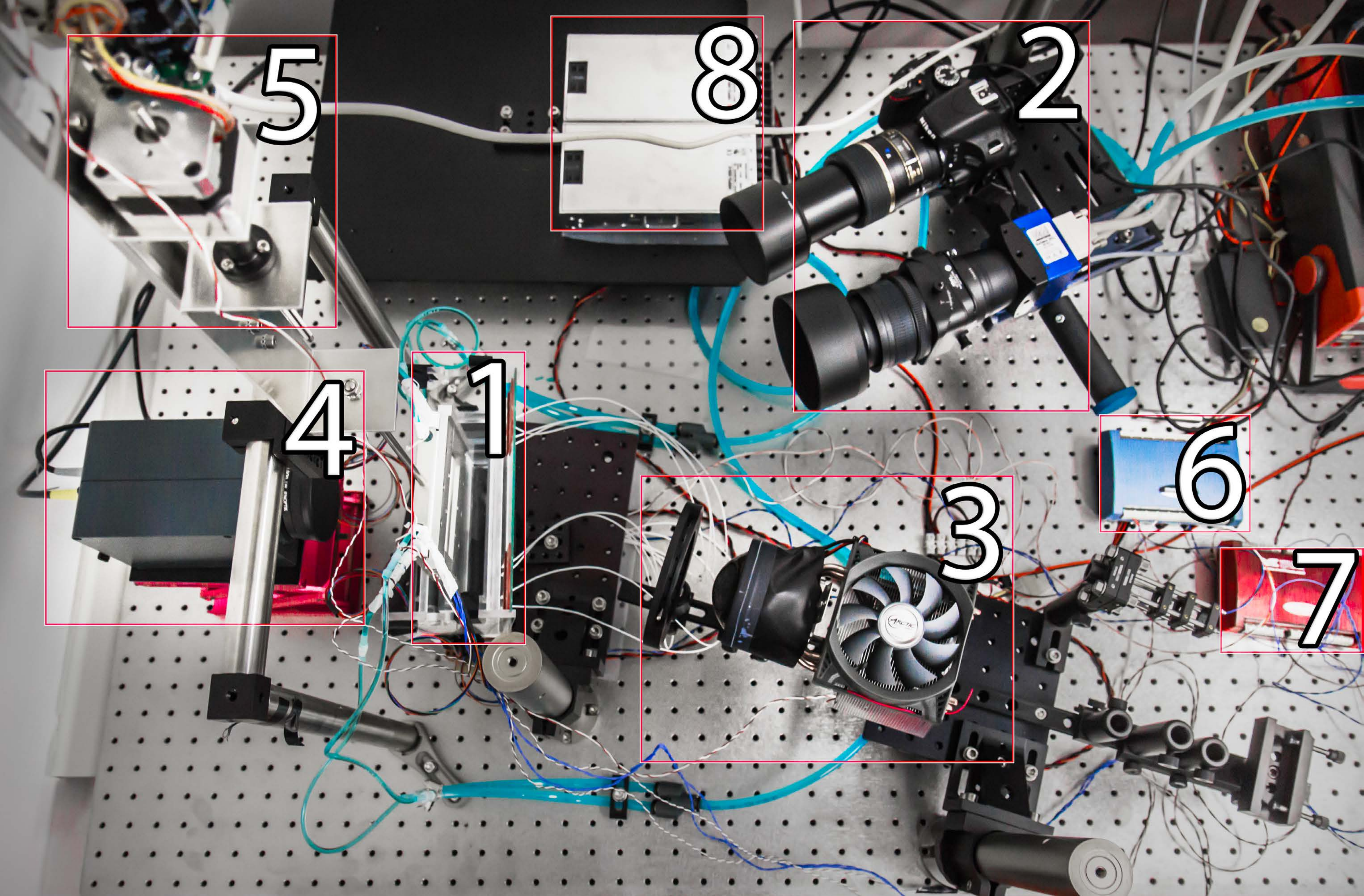}\\
  \caption{Measurement setup - top view. (1) frame assembly, (2) video and still frame cameras, (3) lamp, (4) IR camera, (5) motor assembly, (6) voltage generator, (7) temperature measurement unit, (8) power supplies. Heat exchanger is not shown. top and back lid of the measurement chamber where removed for better visibility.}\label{fig:setupphoto}
\end{figure}

Figure \ref{fig:setupphoto} is a photograph of the complete experimental setup. The details of the system will be described in section \ref{sec:exp_details} and \ref{sec:civ}.
\subsection{Key features}  
\begin{figure}%
\centering
        \begin{subfigure}{0.4\textwidth}
        \def\stackalignment{l}
            \topinset{\large{\bfseries(a)}}{\includegraphics[width=\textwidth]{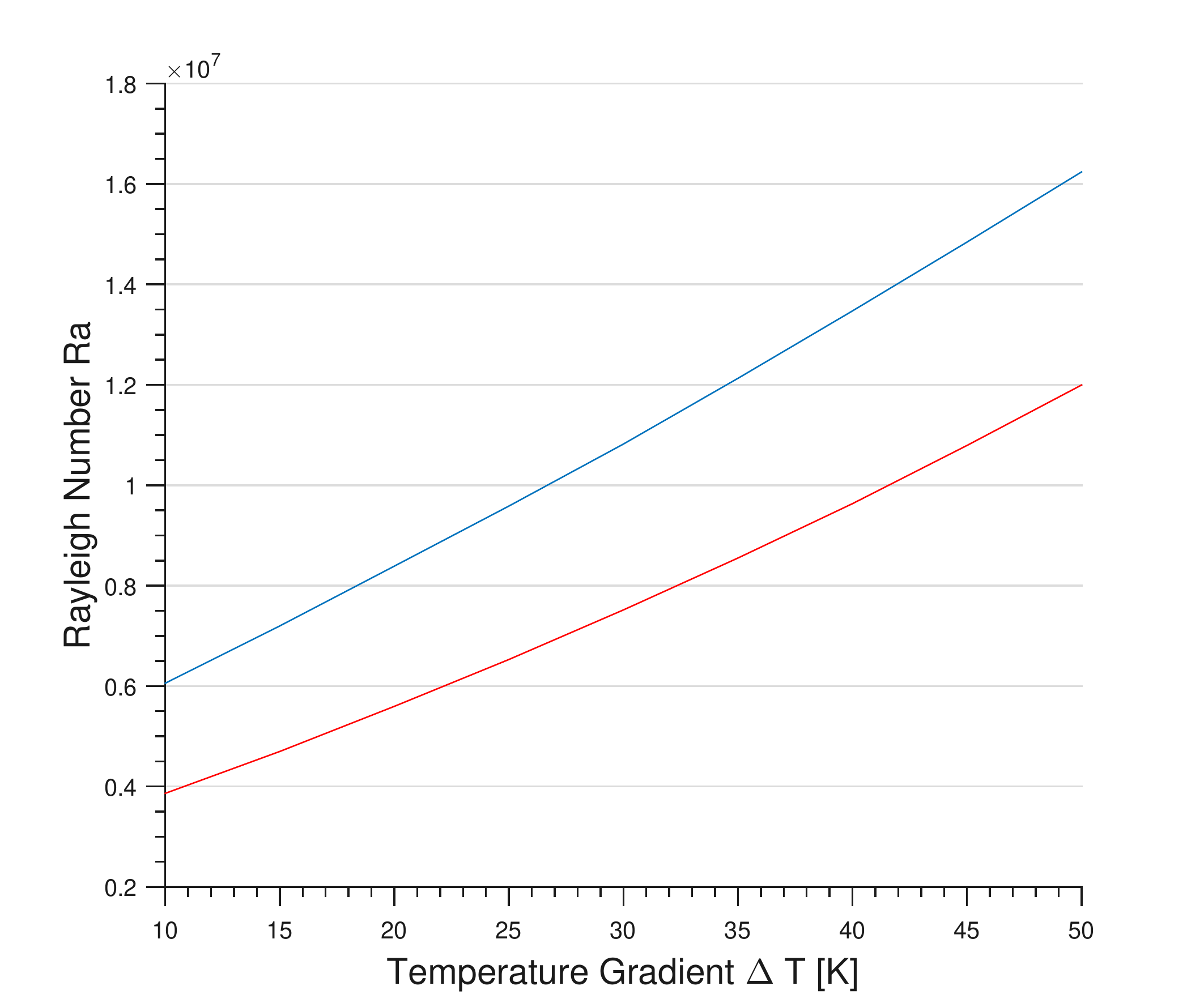}}{.4cm}{1.3cm}
        \end{subfigure}
        \begin{subfigure}{0.4\textwidth}
            \def\stackalignment{l}
            \topinset{\large{\bfseries(b)}}{\includegraphics[width=\textwidth]{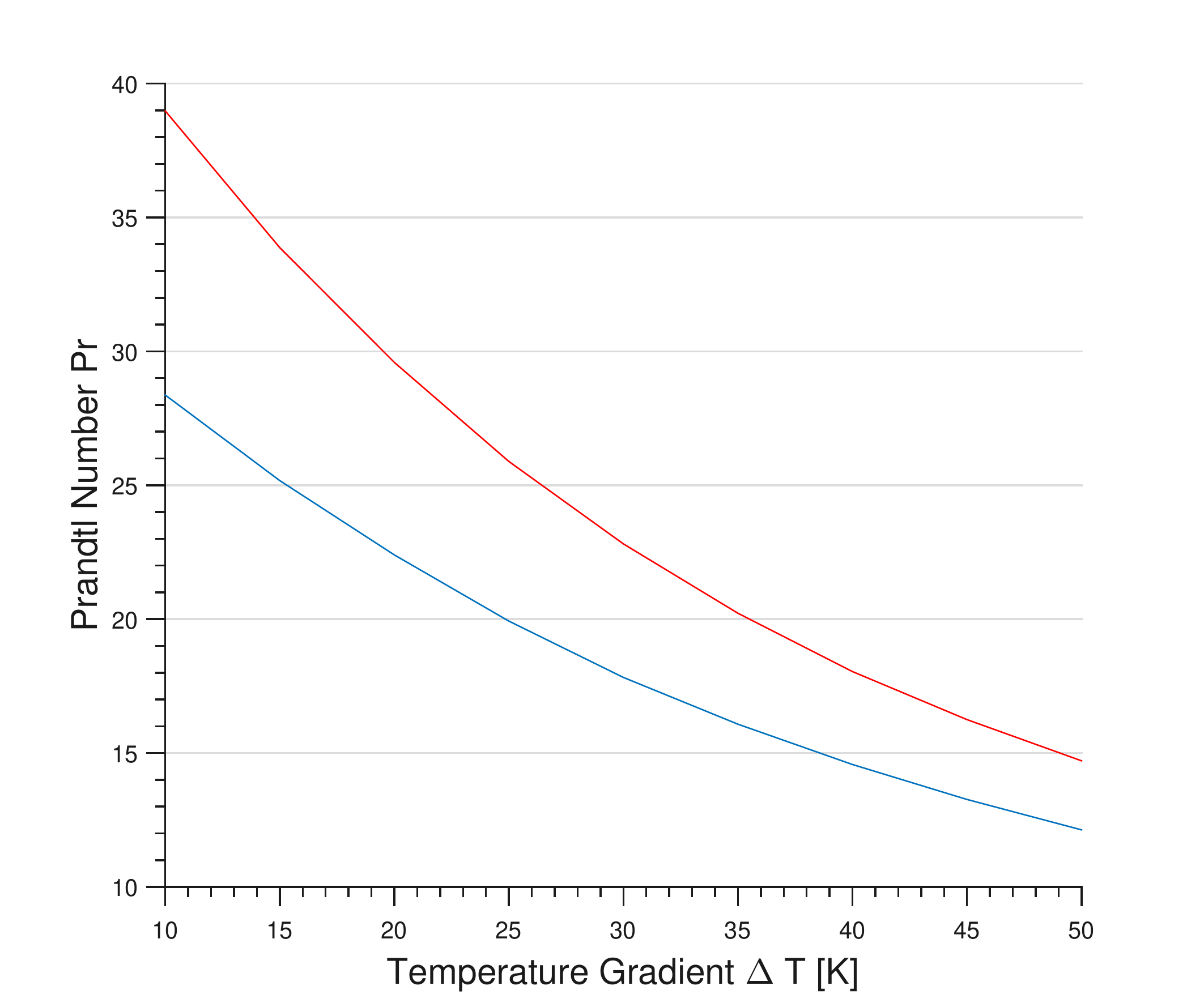}}{.4cm}{1.3cm}             
        \end{subfigure}
    \caption{(a) Rayleigh number and (b) Prandtl number scaling with available temperature and viscosity ranges for $10 \%_{vol}$  (blue) and $25 \%_{vol}$ (red) glycerol.}
    \label{fig:parameters}%
\end{figure}

Our setup is focused on the study of  the influence of the aspect ratio on the convection pattern by a frame, holding the TLF, which can be varied in width and height independently. A closed measurement chamber meets the requirements for controlled ambient parameters. The temperature control units for a temperature gradient similar to classic Rayleigh-Bénard-Convection. The liquid film is heated at the bottom and cooled at the top by an array of thermo-electric couples (TEC) which is built into the frame which supports the freestanding thin liquid film. In table \ref{tab:param} the system parameters are summarized and additionally the feasible Rayleigh and Prandtl numbers are shown in figure \ref{fig:parameters}. The main features of the setup are:

\begin{itemize}
\setlength\itemsep{0em}
  \item     The aspect ratio $\Gamma$ of the film is variable with a  film height range of  $h_{film}= \unit[5 -30]{mm}$ and film width range of $w_{film}= $ \unit[5 -50]{mm}. This covers an aspect ratio of $\Gamma=0.16 - 10$.
  \item The Rayleigh number can be varied in the range of $R_a=$\numrange{e6}{e7} by adjusting the temperature gradient, cf. Fig \ref{fig:parameters}(a).
  \item The Prandtl number can be adjusted by varying the viscosity, respectively glycerin content, cf. Fig. \ref{fig:parameters}(b).
  \item The case is completely transparent to allow optical measurements of the film. The rear panel of the case can be exchanged with a panel containing a Germanium window to allow thermal imaging with an infrared camera.
  \item The front window is angled downwards with respect to the TLF-plane to prevent reflection from the light source off the front glass into the camera to enhance imaging contrast.
  \item The Temperature of the horizontal lower and upper frame parts are precisely controlled by an array of thermoelectric couplers (TECs).
  \item The front window can be heated to avoid condensation and regulate ambient temperature.
  \item The humidity inside the measurement cell is measured and controlled by evaporating water from a separate chamber.
  \item The pressure is set by ambient pressure, but can be controlled by sealing the setup. We do not use this feature, though.
\end{itemize}

\begin{table}
\scalebox{0.8}{
\begin{tabular*}{0.75\textwidth}{@{\extracolsep{\fill} } l c r}
 \hline
  Area                                                          & $A$                          & $0.025 - 1.5 \cdot 10^{-6} \,m^2$ \\
  Density                                                     & $\rho$                   & $1019 - 1080 \,\unitfrac{g}{m^3}$ \\
  Dynamic viscosity                                & $\mu$                    & $0.7 - 3.7 \,\unitfrac{Ns}{m^2}$\\
  Kinematic viscosity                             & $\nu$                     & $0.7 - 3.1 \cdot 10^{-6} \,\unitfrac{m^2}{s}$ \\
  Mean velocity                                        & $u$                         & approx. $2 \cdot 10^{-2}\,\unitfrac{m}{s}$ \\
  Raleigh number                                   & $Ra$                        & $0.37 - 1.62\cdot 10^{7}$ \\
  Prandtl number                                   & $Pr$                        & $12 - 38$ \\
  Specific molar heat capacity            & $c_p$                     & $207 - 254 \,\unitfrac{J}{mol K}$\\
  Surface tension                                    & $\sigma$              & $34 \cdot 10^{-3} \,\unitfrac{N}{m}$ \\
  Thermal diffusivity                               & $\alpha_{th}$    & $5.8 - 8.0 \cdot 10^{-8}\,\unitfrac{m^2}{s}$ \\
  Thermal conductivity                          & $\kappa_{th}$    & $0.49 - 0.60 \,\unitfrac{W}{m \cdot K}$ \\
  Thermal exp. coefficient (vol.)      & $\beta$                 & $255 - 362 \cdot 10^{-6}\,K^{-1}$ \\
   Hamaker constant                              & $C_{H}$                 & $0.5 \cdot 10^{-20} \,J$ \\
 \hline
\end{tabular*}
}
\caption{Parameters of the experiment.}\label{tab:param}
\end{table}

\subsection{Measurement Chamber}
\begin{figure}
  \centering
  \includegraphics[width=.7\textwidth]{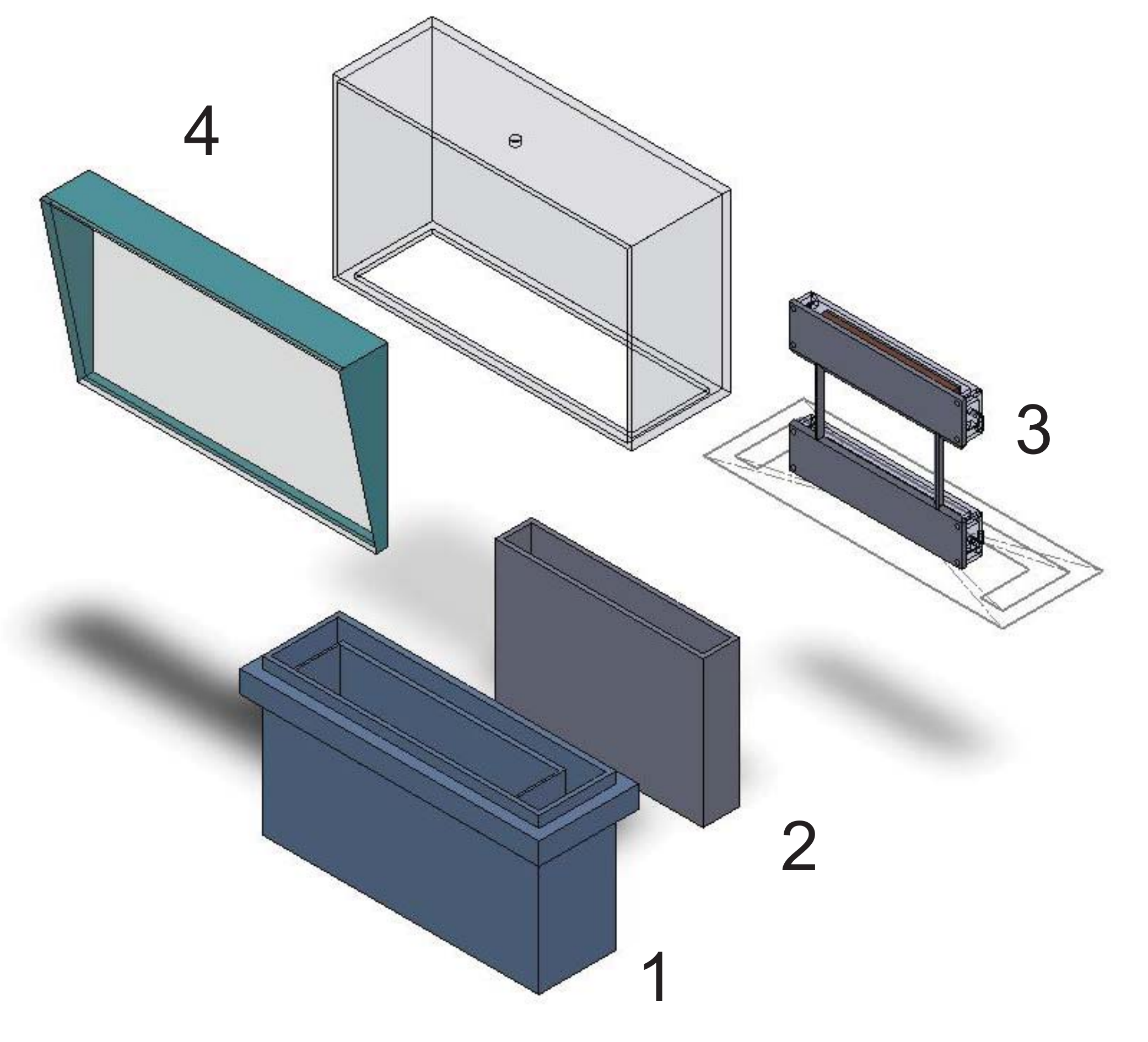}\\
  \caption{Assembly explosion rendering,(1)  container, (2) solution reservoir, (3) frame assembly, (4) top enclosure}\label{fig:assembly_explosion}
\end{figure}

The measurement cell is composed of four parts, displayed in an explosion view rendering in figure \ref{fig:assembly_explosion}. The container (1) has a rectangular cavity which fits the solution reservoir (2). Additionally, there is a U-shaped reservoir in the  container with two resistive heating elements built in. This reservoir is filled with deionized water and used to regulate the humidity inside the chamber via an automated control loop using a humidity sensor. The frame assembly (3) fits into the film solution reservoir (2) in which it can be submerged and pulled out by an automated motorized control. The top enclosure of the setup (4) sits on top of the bottom reservoir and shields the thin liquid film thus allowing for a precisely controlled atmosphere (temperature and humidity) inside the measurement cell.

The dimensions of the measurement cell are based on the frame (3) dimensions \unit[85 - 90 - 25]{mm} (width -- height -- depth). Therefore the outer dimensions of the complete measurement cell are \unit[190 - 170 - 100]{mm} (width -- height -- depth). Furthermore, the cell is modular to allow for easier cleaning as TLFs are very susceptible to contamination and dust particles.

\subsection{Frame Assembly}
\begin{figure}
  \centering
  \includegraphics[width=.5\textwidth]{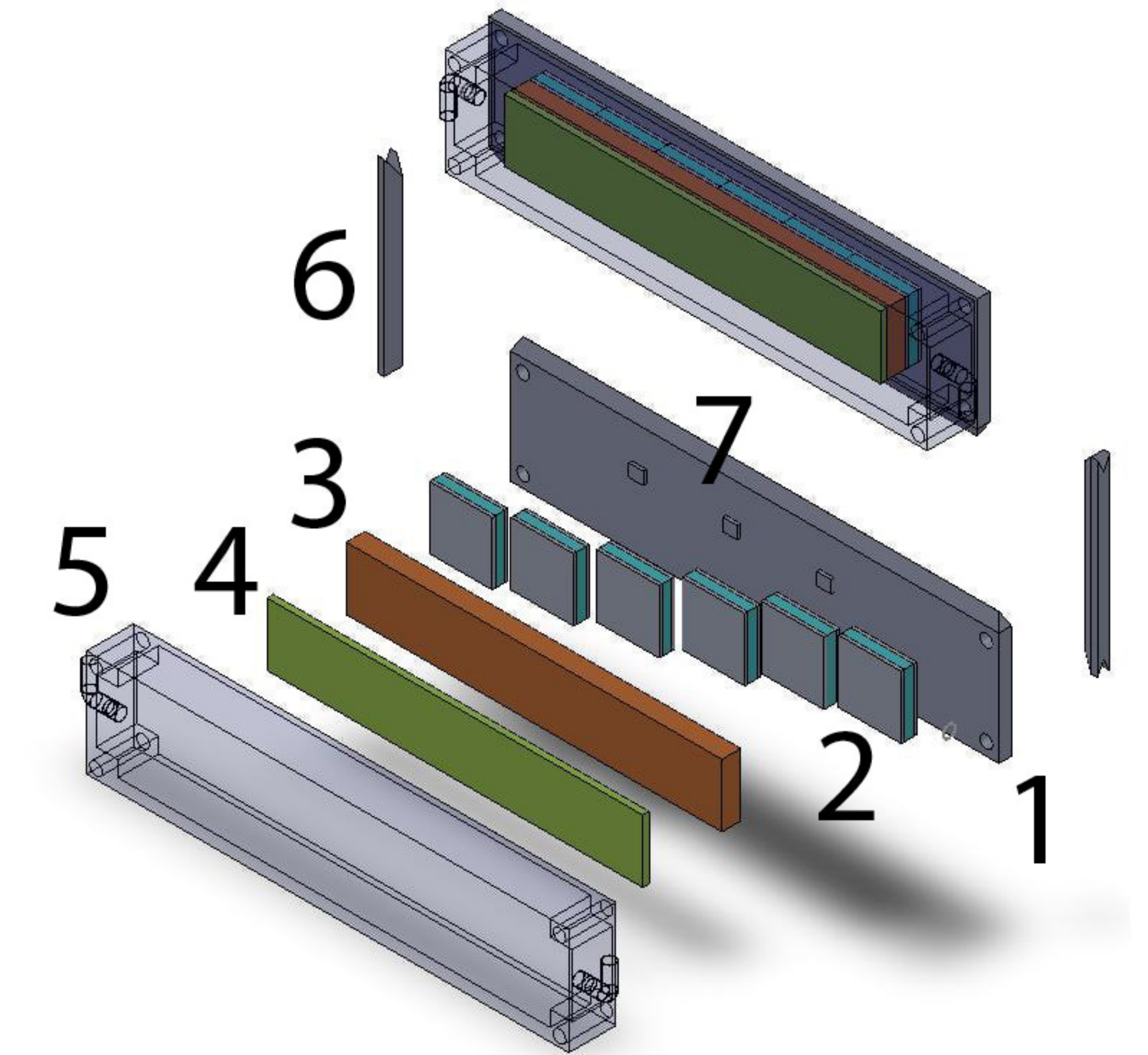}\\
  \caption{frame assembly explosion view. (1) aluminum base plate, (2) thermo-electric couplers array, (3) water cooling block, (4) insulation mat, (5) top cover, (6) adjustable vertical frame parts, (7) temperature sensors.}\label{fig:frame_explosion}
\end{figure}

\begin{figure}
  \centering
  \includegraphics[width=.5\textwidth]{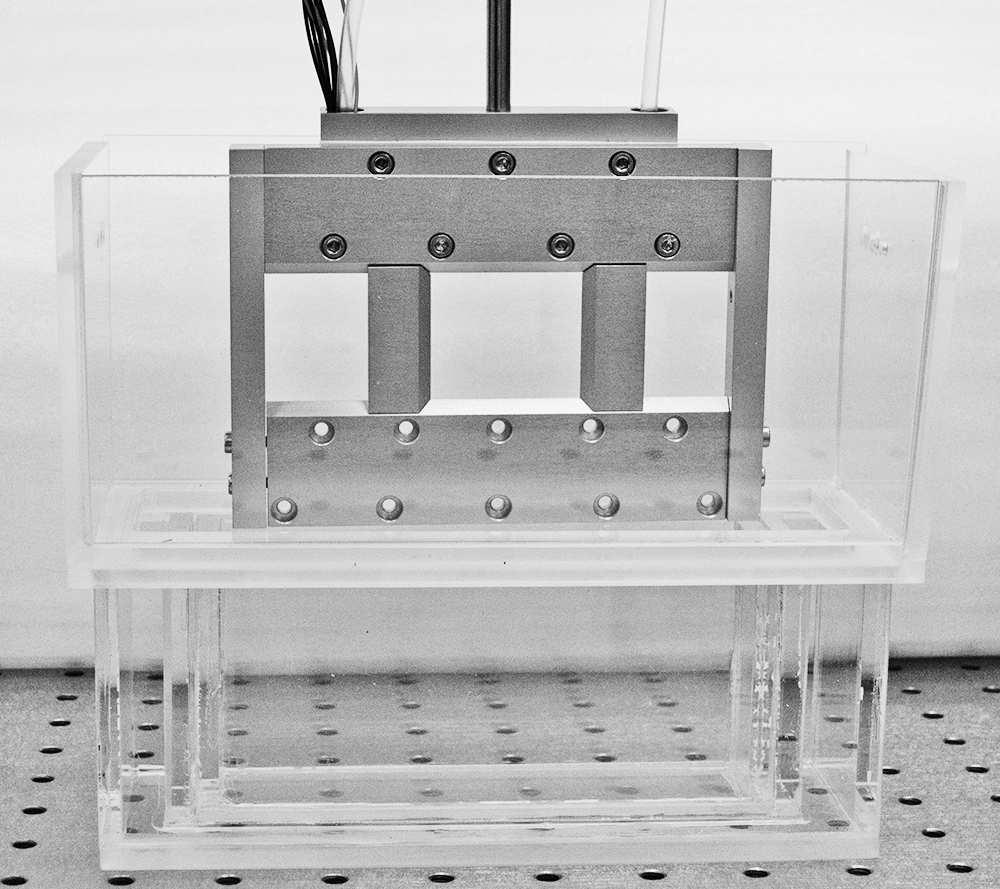}\\
  \caption{frame inside the measuring chamber - front view.}\label{fig:framephoto}
\end{figure}

The frame is a modular assembly of four edge elements and two temperature control units, which can be mounted at the top and bottom. The edges which are holding the thin liquid film are sharply angled to allow a formation of the thin liquid film at the center position. Figure \ref{fig:framephoto} shows the position of the frame inside the measuring chamber with the attached temperature control device at the top and at the bottom and figure \ref{fig:frame_explosion} displays an explosion rendering of these devices. The base aluminum plate (1) is heated or cooled by an array of thermo-electric couplers (TEC) (2) which in turn are cooled by a separate closed water cooling loop via the copper cooling block (3). The assembly is then closed with a top cover (4) and a sealing to make it waterproof. Cabling and the outlets of the water cooling loop exit through sealed tubes at the top of the temperature control units. To vary the aspect ratio of the TLF the vertical frame elements can be continuously shifted along the  horizontal frame elements. The height of the frame can be adjusted inserting vertical frame elements of different length. Figure \ref{fig:frame_explosion} displays an explosion rendering of the frame assembly

\section{Experimental Details}
\label{sec:exp_details}
\subsection{Thin Film Generation}
Macroscopic foam films are obtained by spanning an aluminum frame with surfactant solution, creating a quasi two-dimensional planar liquid. To create a TLF the frame assembly is submerged into the solution reservoir and by pulling the frame out liquid gets trapped between the edges of the frame, much like creating a soap bubble. The frame motion is controlled by a stepper motor which enables jerk-free motion to avoid any shock to the frame assembly which could cause the film to rupture. The speed of this process determines the initial thickness of the thin liquid film. The film then forms a wedge-like profile due to gravity. After the initial draining phase of excess fluid the gravitationally driven thinning process is orders of magnitude slower than the fluid transport by heat convection. Essentially, this enables us to keep the amount of liquid inside the frame constant as the liquid is transported as no drainage over the bottom meniscus occurs. Any loss in fluid volume can be attributed to the evaporation at the bottom heating element due to the lower concentration of water in the air.

The amount of fluid in the frame can be estimated by the number of visible stripe patterns of the captured reflection image. Details concerning the data acquisition are discussed in section \ref{sec:civ}.

\subsection{System Control}
\begin{figure}
  \centering
  \includegraphics[width=\textwidth]{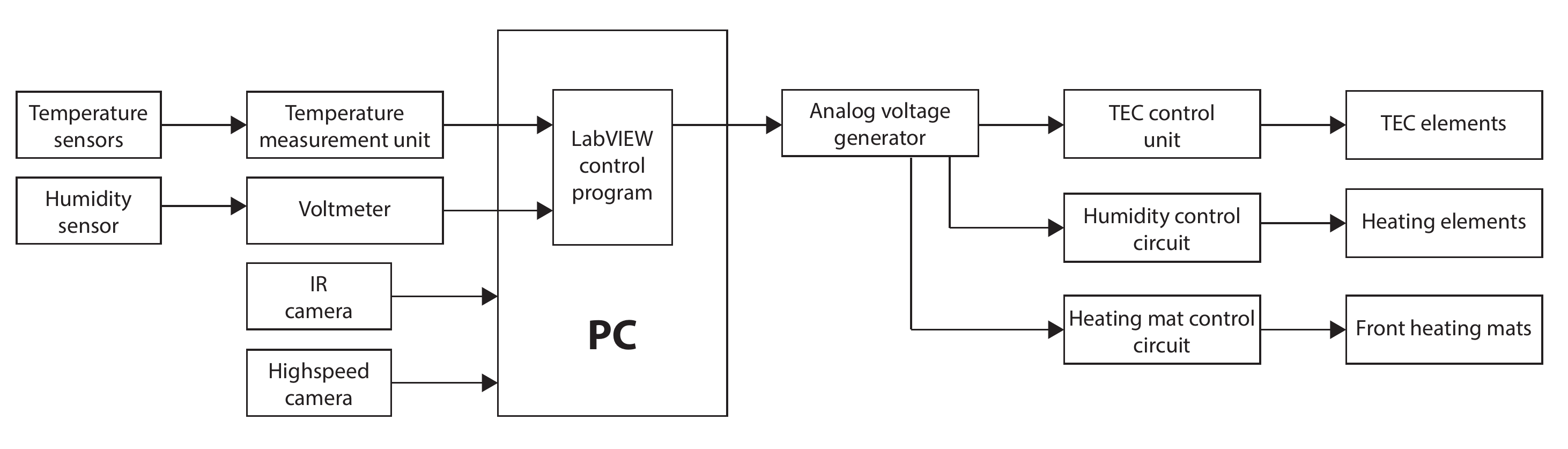}\\
  \caption{Schematic of the overall setup control}\label{fig:schematic}
\end{figure}

The overall sensory input and control circuits are sketched in figure \ref{fig:schematic}. The main components are the temperature input and analog voltage output which are managed by the LabVIEW application.
For the most part there exist two temperature control schemes in research and industry: either the devices is connected to a large thermal mass whose temperature is regulated by a flow-through cooler or a heat pump in form of a thermoelectric cooler (TEC) is employed. The type of device implemented depends on the scale of the setup, the expected temperature variation and the desired speed of temperature changes ($\unitfrac{dT}{s}$). The dimensions of our setup are relatively small and the thermal mass of the frame elements is low which favours a temperature control setup based on TEC units which operate on a high update frequency to insure optimal temperature balance. Additionally, this enables us to quickly change the temperature with a controlled ramp up.

The TEC-arrays in the top and bottom frame unit are connected to a control unit which adjusts the amount of current flowing through the array to achieve a precisely controlled temperature. The bottom temperature control unit contains 6 TEC modules and the top unit contains 5 modules. Each module covers an area of \SI{1}{\cm^2} and can pump a maximum of $Q_c=\SI{5}{\watt}$  allowing rapid temperature changes. The generated excess energy is dissipated by the water-cooled copper cooling blocks, which are connected to a large external heat exchanger. The wiring allows to control each TEC element individually to achieve better homogeneity of the temperature profile. In the current setup the elements for each section are wired in series and monitored by 4 temperature sensors.

 The PID control loop for top and bottom temperature and humidity are realized in LabView together with the motor control for the automated frame submersion into the solution reservoir.  The necessary wires and tubing for the water cooling loop exit the measurement chamber through holes in the side walls.

 Temperature gradients up to $\Delta T = \SI{40}{\kelvin}$ are possible. The maximum temperature difference is limited by decreased film stability at high temperatures and increased condensation at the top border of the frame at temperatures close to the freezing point of the solution.

 The humidity in the chamber is controlled by evaporating water from the water reservoir using two heating cartridges. Ideally, the concentration of water in the air inside the chamber is close to the dew point for the respective ambient temperature to avoid any evaporation from the TLF. To prevent the generated water vapor to condense on the observation window an array of heating mats is used. This system can also be used to adjust the ambient temperature inside the cell.

\subsection{Surfactant Solution}
The base solvent is deionized water, as it reduces contaminations with
non-conducting carbon chain remnants or residual ion concentrations
\cite{muruganathan2004foam}.

Glycerin is also required to achieve macroscopically large films with an area of
several square centimeters or meters by adjusting the bulk viscosity \cite{martin1998spectra,kellay1995experiments}.
Due to its high viscosity ($\eta=945~{\mathrm{ mPa\,s}^{-1}}$, compared to $\sim
1~{\mathrm{ mPa\,s}^{-1}}$ for water) it reduces the thinning speed of vertical films
and allows the film to remain stable for a longer period of time
\cite{isenberg1992science}. 
Typically, a concentration of 25~$\%_{\mathrm{vol}}$ is used, yielding a net
viscosity of $\sim 2~{\mathrm{ mPa\,s}^{-1}}$ , cf. \cite{rutgers2001conducting}.

The electrolyte concentration has a major effect on film
stability and thickness \cite{zhang1996electrolyte}. The addition of
ions shields the electrostatic double-layer repulsion as described by
the DLVO (Derjaguin-Landau-Verwey-Overbeek) theory \cite{derjaguin1989theory}
hence permitting the appearance of a black film.  Here, sodium chloride was
chosen, since
it has negligible influence on surface tension and critical micelle
concentration (CMC). To eliminate any surface active contaminations,
the laboratory grade $NaCl$ was roasted at \SI{600}{\celsius}.

The chosen surfactant, n-dodecyl-$\beta$-maltoside ($\beta\mathrm{-}C_{12}G_2$), belongs to a new generation of
sugar-based surfactants which are non-toxic, biodegradable and of low
cost. $\beta\mathrm{-}C_{12}G_2$ is a non-ionic surfactant which consists
of hydrophilic head group, made up of two glucose rings ($C_6H_{12}O_6$)
connected by an ether bond, and a hydrophobic
alkyl chain ($C_{12}H_{25}$). The surfactant is soluble in water so
that the hydrophilic head group is submerged in the core liquid. The
hydrophobic tail is oriented normal to the liquid-air-interface and is
not in contact with film surface.  All experiments were performed with
a concentration above the CMC, to guarantee that the equilibrium
surface tension remains constant. Above CMC, deviations in surface
surfactant density are compensated by diffusion of surfactants from or
to the bulk liquid.

\section{\label{sec:civ} Data Acquisition and Processing}
A non-invasive technique, Color Imaging Velocimetry (CIV), is used to capture
and analyze the turbulent motion \cite{winkler2013mixing,winkler2013exponentially}.
In classic PIV (particle imaging velocimetry) applications small particles are seeded and traced to calculate flow properties. This technique has been previously employed in the study of soap films
\cite{cheung1996diffusion,vorobieff1999soap,vorobieff2001imaging}.
However, for our setup the use of conventional PIV particles is not possible due to the fragility of the film and the feedback of the particles on
the flow: the film thickness one can still observe with CIV using visible light lies at \SIrange{0.1}{2}{\micro\metre} whereas the particles have a radius of \SIrange{0.1}{5}{\micro\metre}. 

\subsection{Color Imaging Velocimetry}

CIV is based on the thickness mapping of the TLF by
capturing the reflection of a diffuse light from the surface
\cite{atkins2010investigating,Joosten-88}. The interference pattern
can be recorded monochromatically or in full color
\cite{princen1965optical,gharib1999visualization,yang2001interpretation}. It
is mostly used as a relative measure of the thickness distribution or
an immediately available visualization of the flow structures. In order to get
quantitative results a careful design is needed \cite{winkler2011droplet}.

The interference of incident and reflected light yields
a striped pattern, which can be used to infer the film thickness.
Each color cycle corresponds to the multiples $n$ of the smallest negative interference condition
$(2\cdot n+1) \lambda \eta = 4 h \cos \Theta$, where the refraction index, $\eta$, is
assumed to be temperature-independent; $\Theta$ is the angle of incidence \cite{atkins2010investigating}.

The film is illuminated through this viewing window by an array of
high power white LEDs which are placed behind a diffuser.  Additionally, to avoid direct reflections of the light source from the viewing window, the front observation window is tilted downwards with respect to the liquid film
plane.

\begin{figure}%
\flushleft
        \begin{subfigure}{0.32\textwidth}
        \color{white}
         \def\stackalignment{l}
            \topinset{\large{\bfseries(a)}}{\includegraphics[height=4cm]{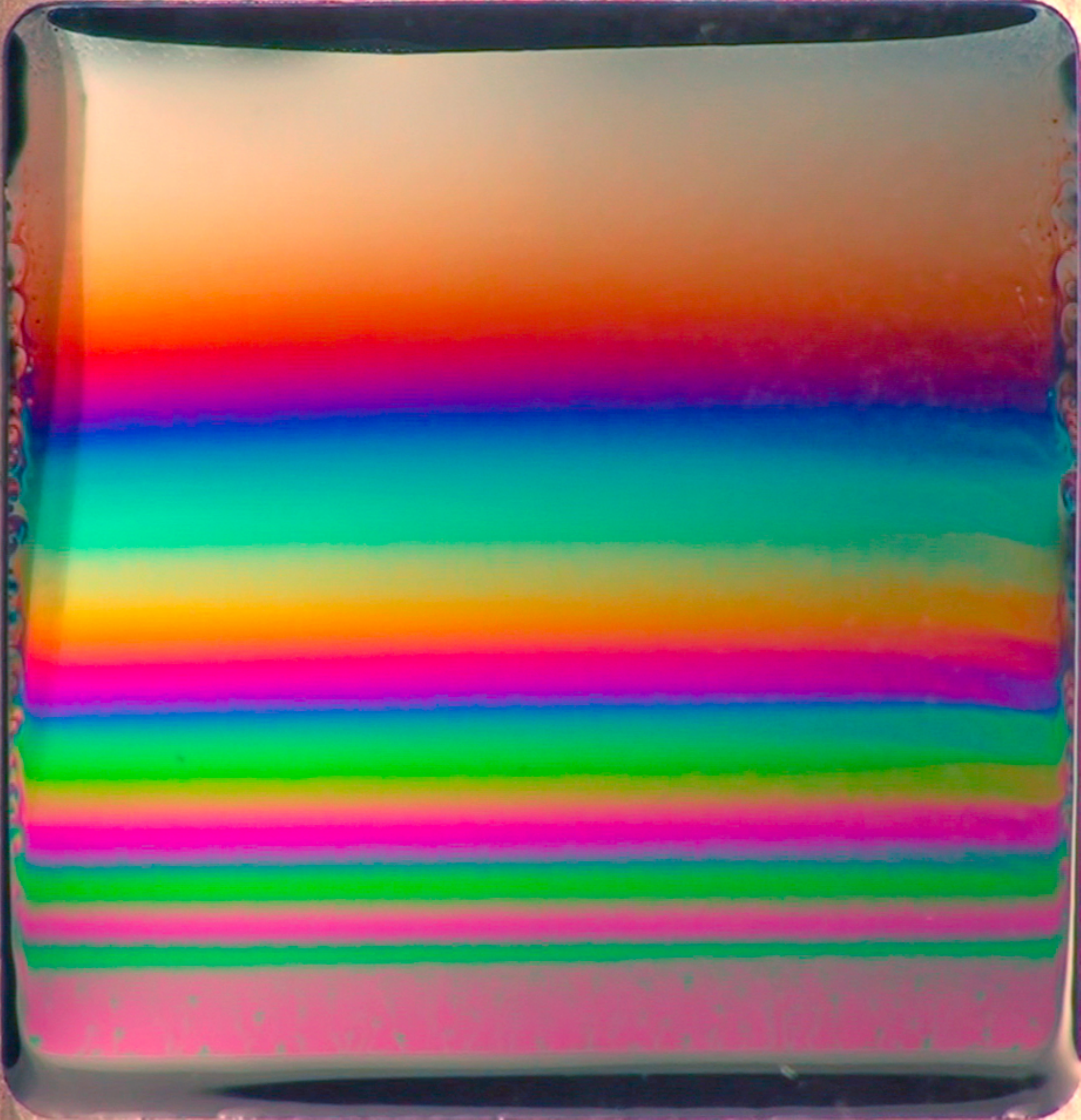}}{.4cm}{.4cm} 
        \end{subfigure}
        \begin{subfigure}{0.32\textwidth}
            \color{white}
            \def\stackalignment{l}
            \topinset{\large{\bfseries(b)}}{\includegraphics[height=4cm]{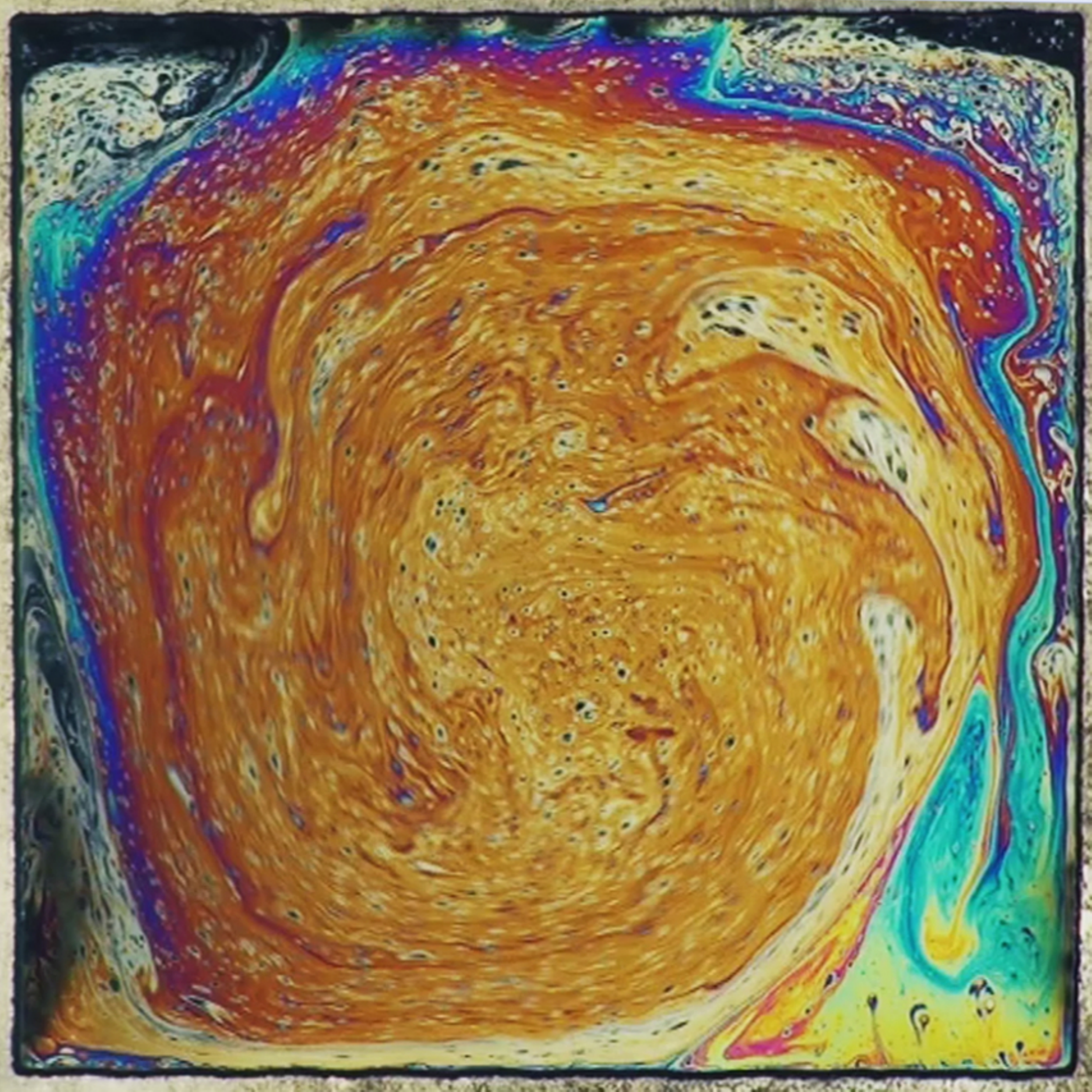}}{.4cm}{.4cm}            
        \end{subfigure}
        \begin{subfigure}{0.32\textwidth}
            \def\stackalignment{l}
            \topinset{\large{\bfseries(c)}}{\includegraphics[height=4cm,width=5cm]{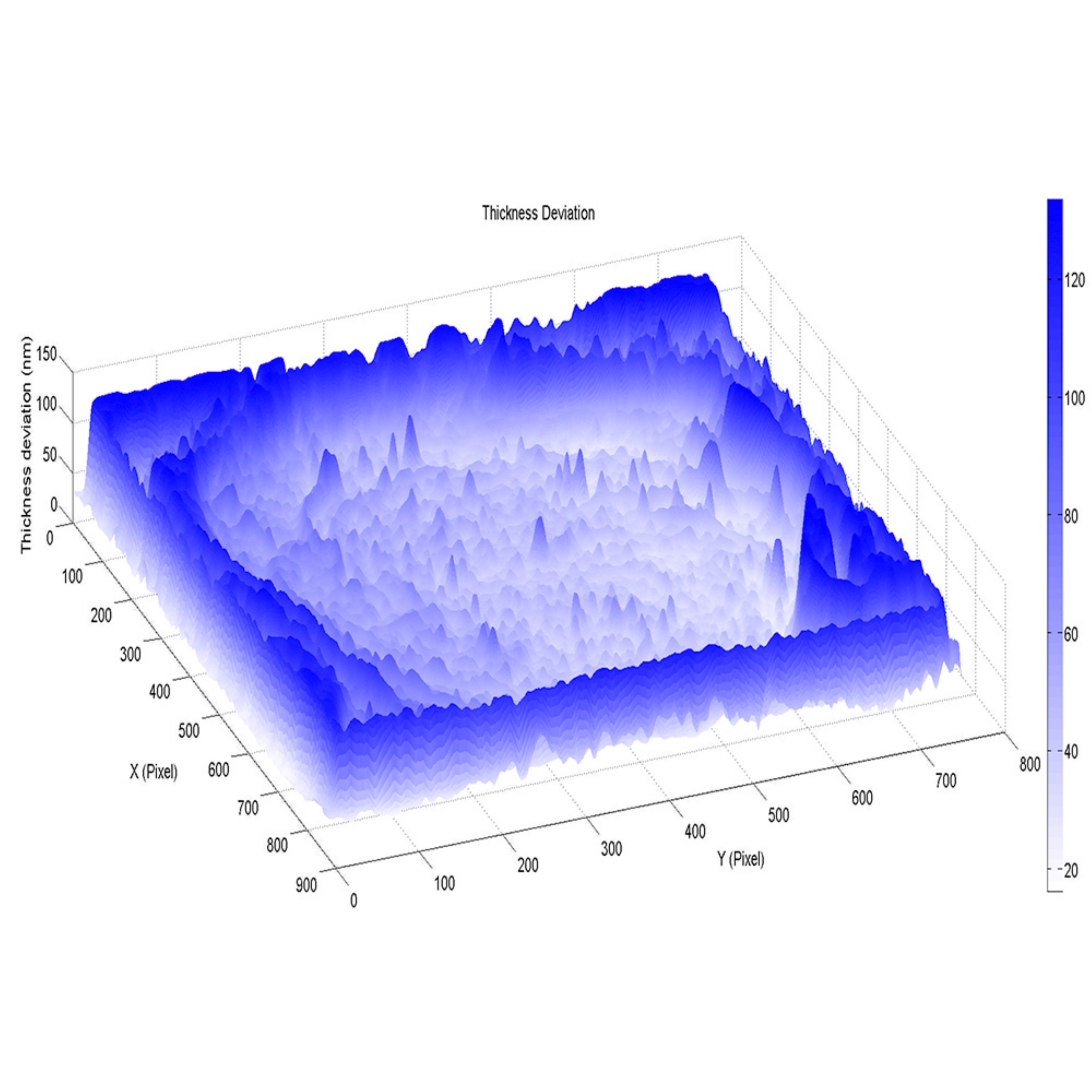}}{.4cm}{.4cm}
        \end{subfigure}
    \caption{Snapshots and thickness profile of the TLF. (a): Stratified TLF with no temperature gradient applied. (b): Fully developed large scale circulation in a TLF. (c): Thickness deviation profile obtained by color analysis (spatial unit: Pixels).}
    \label{fig:snapshots}%
\end{figure}

The diffused light is reflected by the front and back side of the film whereas
the front reflection is shifted by $\nicefrac{\pi}{2}$ due to reflecting of an optically
thicker medium. If the film thickness is much smaller than $\nicefrac{\lambda}{4}$ of the
smallest emitted wavelength the light waves interfere $\nicefrac{\pi}{2}$  out-of-phase and
no light is reflected. Hence the name black film although more precisely it is
transparent. The reason why there is no smooth transition in the intensity of
the reflected light is the abrupt change in thickness when the black film is
formed. At $\nicefrac{\lambda}{4}$ of the smallest wavelength, which is blue ($\sim$\SI{450}{\nano\metre})
in the case of the used white LEDs, the reflected waves
interfere constructively. In addition, every other wavelength is reflected as
well which sums up to a white color with a blue tint. Before that thickness is
reached the film exhibits an orange tint as the film is approximately \SI{200}{\nano\metre}
thick which corresponds to the cancellation of blue light, cf. figure \ref{fig:snapshots}(a).
With this information the recorded color translates into a thickness profile of the TLF \ref{fig:snapshots}(c)
To get an absolute value of the thickness corresponding to a certain
color displayed one needs to take into account that the refractive index of the
solution is slightly higher than that of pure water. At the end of this section, we would like to note that for Newton Black film, the film is transparent and CIV with visible light is not useable.

\subsection{Data Acquisition and Processing}
The reflected light is recorded with a high-speed camera at 500 frames per second with a resolution of \SI{1.3}{\mega Pixel}. As the reflected light cannot be captured at a \SI{90}{\degree} angle of incident, a Tilt-and-Shift lens is used which allows to tilt the focal plane to be parallel to the thin liquid film.
The captured video data are filtered and converted into a binary image to analyze the behavior of domains of the same thickness.

Subsequently, in each frame, all clusters of the same thickness are numbered and
consecutively linked through the following video frames. This enables us to
track the volume, velocity, deformation rate and angular velocity of the moving
fluid filaments. For each cluster the velocity is calculated using the shift of its center of
mass per frame. The deformation rate is given by the scaling of the principal components. Similarly the angular velocity
is given by the rotation of the principal component. Time-averaging over all frames
then delivers the spatial characteristics of the flow field.

\subsection{IR Imaging}
\label{sec:ir_imaging}
The temperature distribution of the thin liquid film is measured non-invasively with the uncooled microbolometer-focal-plane-array IR Camera InfraTec\circledR VarioCAM hr 680.  The key specifications are the resolution of 640x480 Pixel, a capture frequency of up to \SI{60}{\hertz}, temperature resolution of up to \SI{0.03}{\kelvin} and a spectral range of \SIrange{7.5}{14}{\milli\metre}.

Infrared imaging of liquid thin films is non-trivial, as the radiation
intensity emitted from the film is very
low relative to the background noise. A non-uniform thickness can
distort the measurement \cite{minkina2009infrared}. Previously, infrared
imaging was used to gather information about the instantaneous thickness
profile and fluctuations of \emph{strongly stratified} thin liquid films
\cite{zhang2006thermal,zhang2009persistent}. To obtain the actual temperature distribution inside the film it is necessary to achieve a uniform thickness. In contrast to a stratified system the turbulence in our system is vigorous enough to break the stratification and involve the whole thin film in the convective motion. This enables a more homogeneous thickness throughout the film which makes it possible to retain temperature information without differing thermal emissivity due to large variations in thickness. As the IR-window is built into the back of the measurement chamber one can simultaneously observe the thin liquid film optically via the front window.
 Thus the infrared image can be synched to the reflective image and compensated by the thickness profile obtained from CIV to only retain temperature information. Result measurement using this method are work in progress and subject to future research.

\section{Temperature Stability}

\begin{figure}[ht]
  \centering
  \includegraphics[width=.7\textwidth]{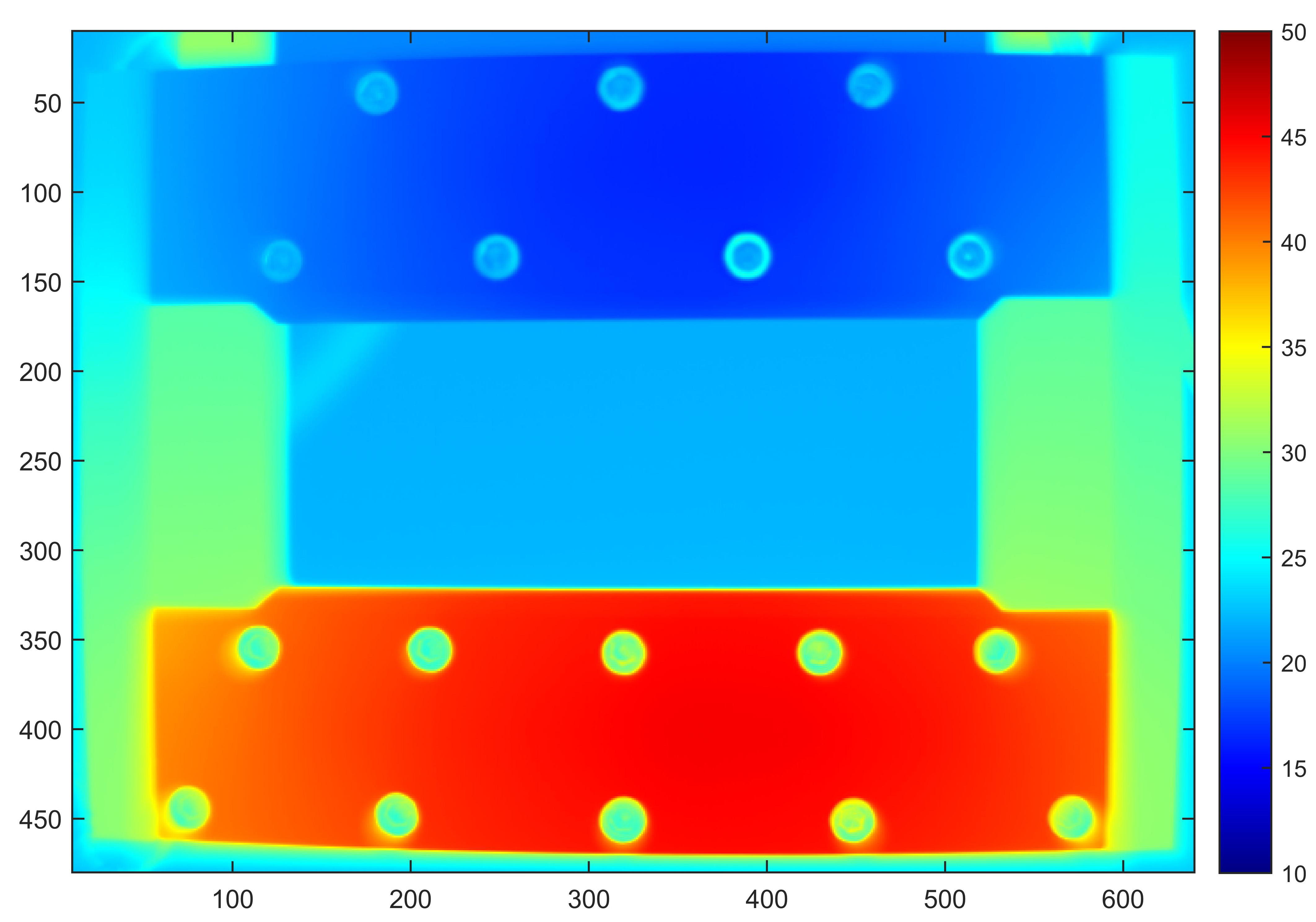}
  \caption{Infrared snapshot of the frame assembly at equilibrium temperature setpoint: $T_{top} =$\SI{18}{\celsius}, $T_{bot} =$\SI{43}{\celsius}. Spatial unit: pixels, temperature in \si{\celsius} }\label{fig:IR_frame}
\end{figure}

\begin{figure}[ht]%
\flushleft
        \begin{subfigure}{0.49\textwidth}
            \def\stackalignment{l}
            \topinset{\large{\bfseries(a)}}{\includegraphics[height=7.5cm]{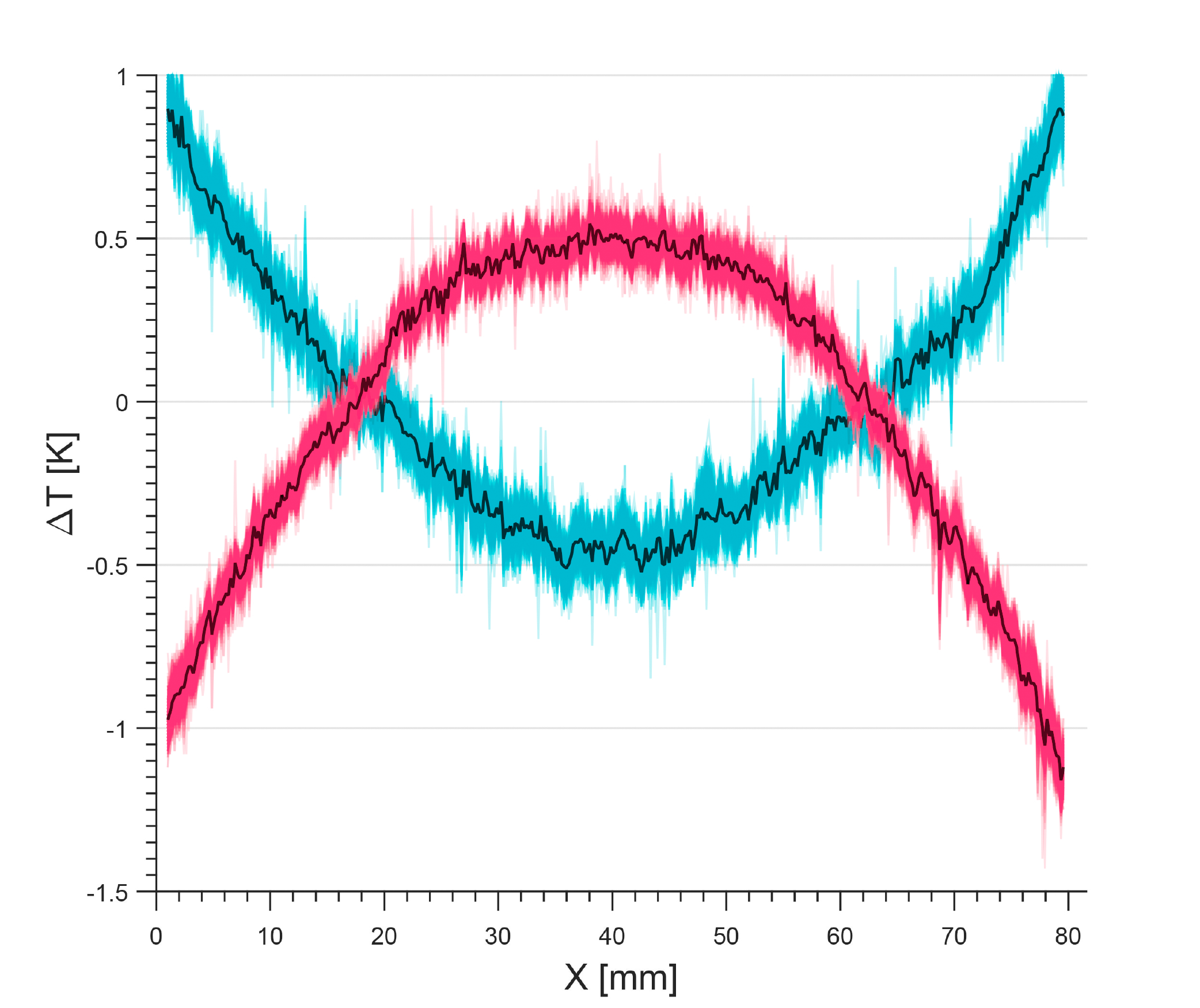}}{1cm}{2cm}                                \end{subfigure}
        \begin{subfigure}{0.49\textwidth}
            \def\stackalignment{l}
            \topinset{\large{\bfseries(b)}}{\includegraphics[height=7.5cm]{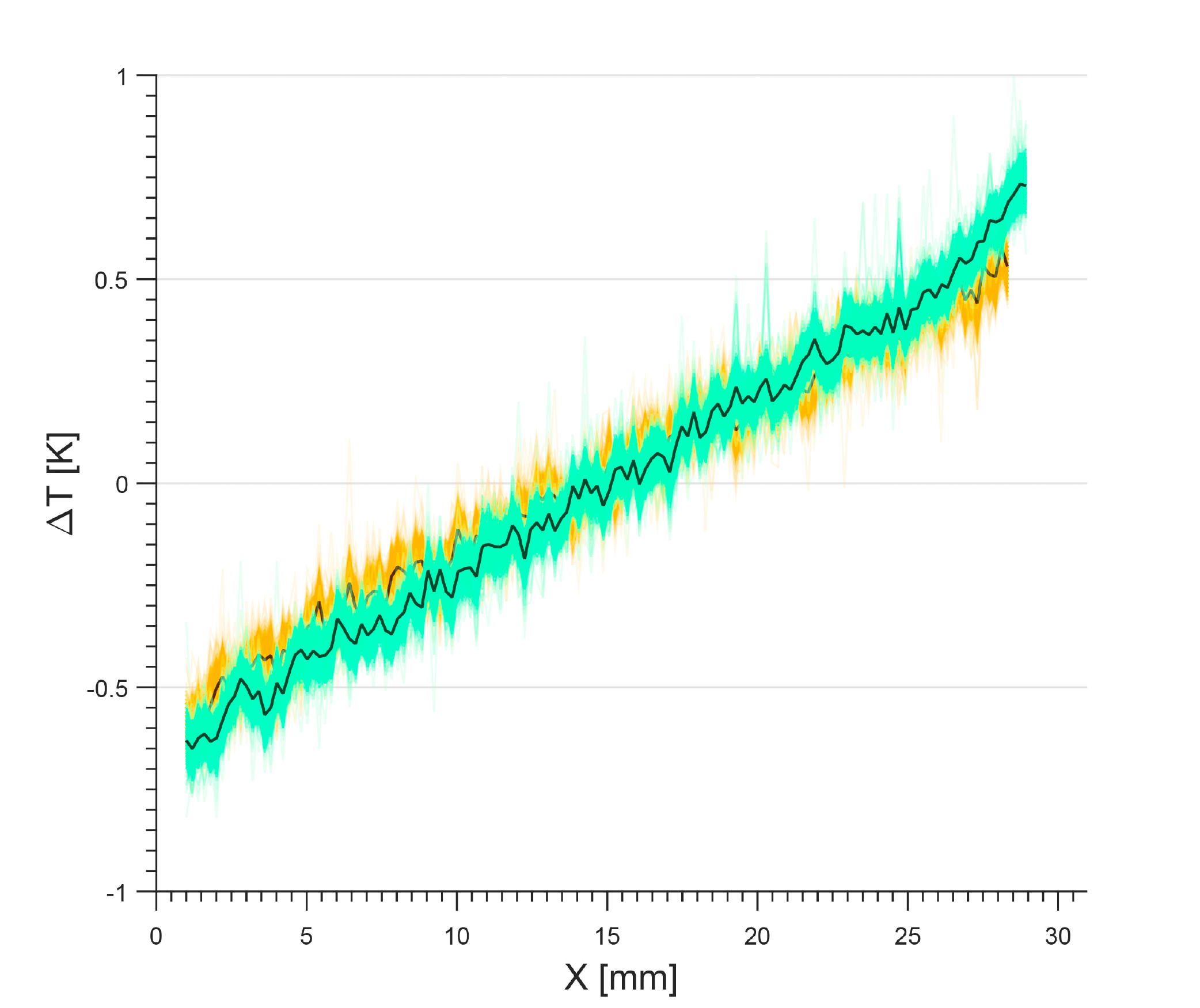}}{1cm}{2cm}
        \end{subfigure}
    \caption{Temperature profiles of frame elements. Black line represents the average.(a): Top element (blue, cooling), and bottom element (red, heating). Variation of the heating element is higher due to the increased temperature difference with the surrounding atmosphere. (b): Vertical frame elements (left and right). Minimal variations in the profile occur due to the slight asymmetry of the bottom and top heating elements.}%
    \label{fig:temp_profile}%
\end{figure}

In this section we will summarize the measured characteristics of the setup. A general overview of the temperature distribution in the frame is given by the infrared image of the frame assembly in figure \ref{fig:IR_frame}. We define the temperature variation $\Delta T_V$ as the deviation from the set temperature along a line profile at the inner edges of the frame elements. The line profiles were measured at \SI{5}{\s} intervals for 30 minutes using the infrared camera, cf. Fig. \ref{fig:temp_profile}.
The setup was left to equilibrate for 20 minutes prior to the start of the measurement. The acquired temperature profiles were averaged for different temperature gradients $\Delta T$ and show a maximum temperature variation of $\Delta T_{V,max}=$\SI{1.5}{\kelvin}. This may seem rather high in comparison to existing, larger volume Rayleigh-Bénard cells. However, previously published experimental designs aimed at thermal convection in thin liquid films do not elaborate on the temperature variation at all \cite{seychelles2008thermal,martin1998double,zhang2005velocity}. The main difficulty in achieving a lower $\Delta T_V$ is the low thermal mass of the frame elements and the control of the TEC elements as a single array. For frame width of \SI{20}{\mm} and below we achieve a temperature variation of $\Delta T_V=$\SI{0.2}{\kelvin}.

\begin{figure}[ht]%
\flushleft
        \begin{subfigure}{0.32\textwidth}
            \def\stackalignment{l}
            \topinset{\large{\bfseries(a)}}{\includegraphics[height=4.5cm]{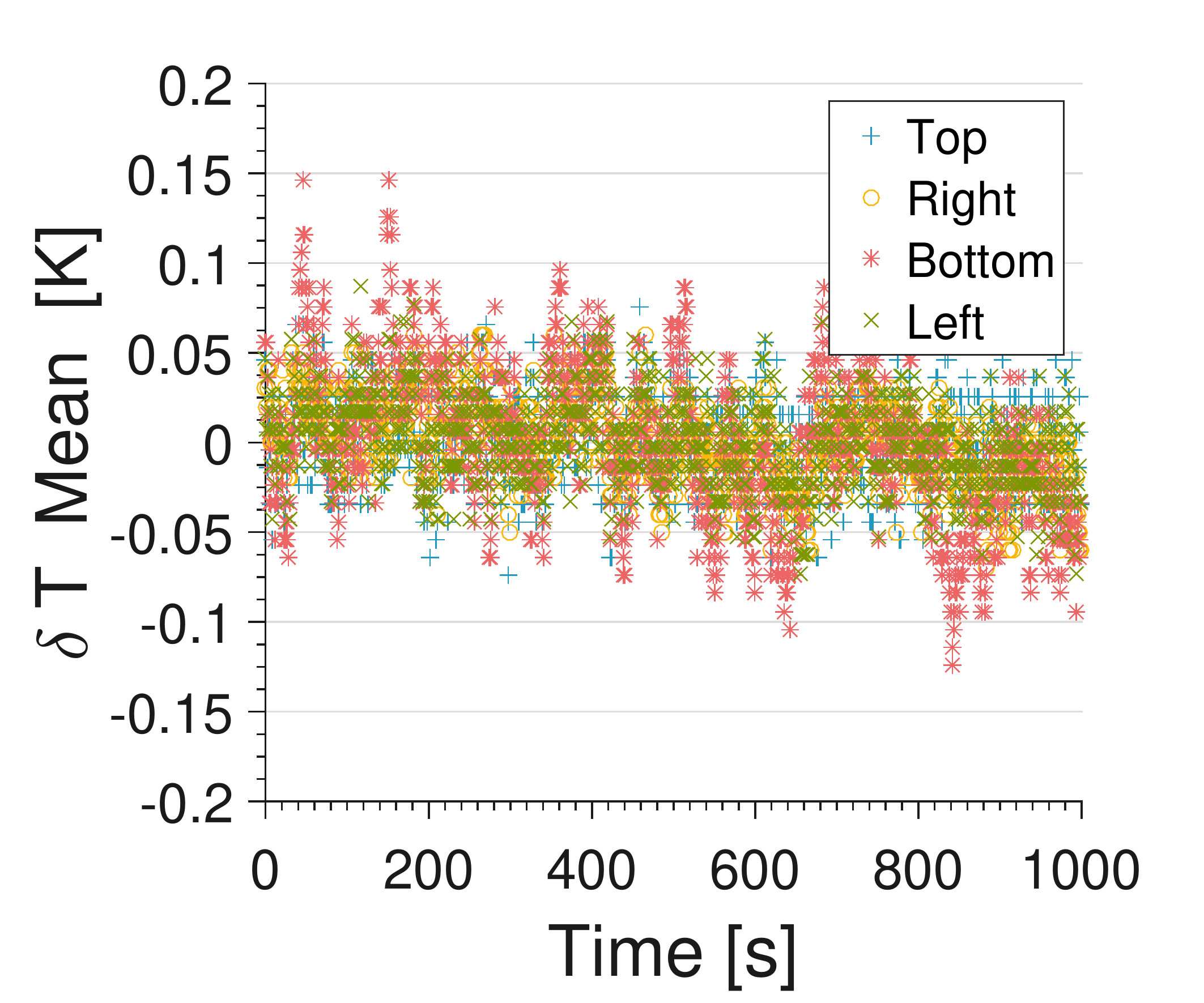}}{0.4cm}{1.4cm}
        \end{subfigure}
        \begin{subfigure}{0.32\textwidth}
        \def\stackalignment{l}
            \topinset{\large{\bfseries(b)}}{\includegraphics[height=4.5cm]{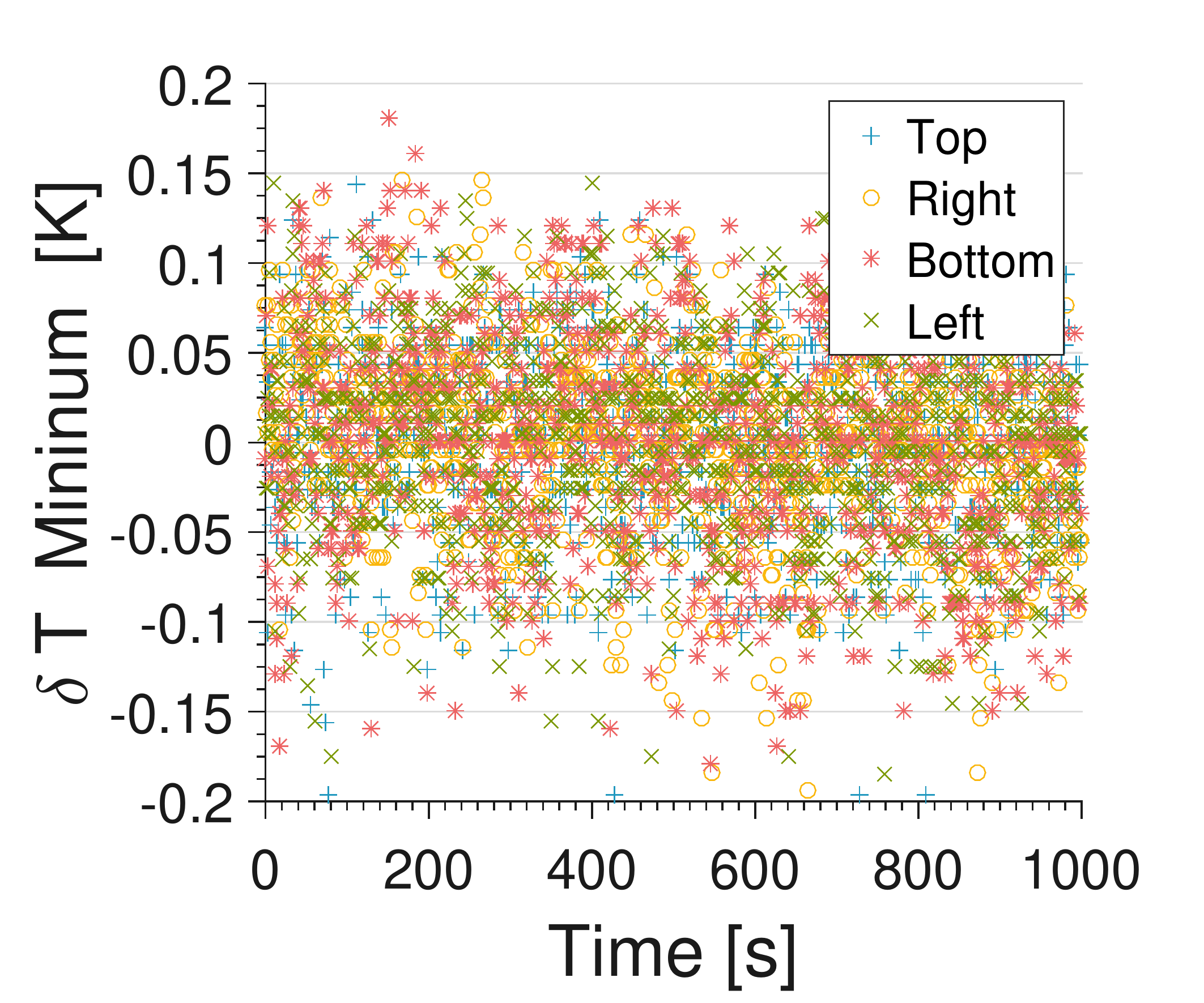}}{0.4cm}{1.4cm}
        \end{subfigure}
        \begin{subfigure}{0.32\textwidth}
        \def\stackalignment{l}
            \topinset{\large{\bfseries(c)}}{\includegraphics[height=4.5cm]{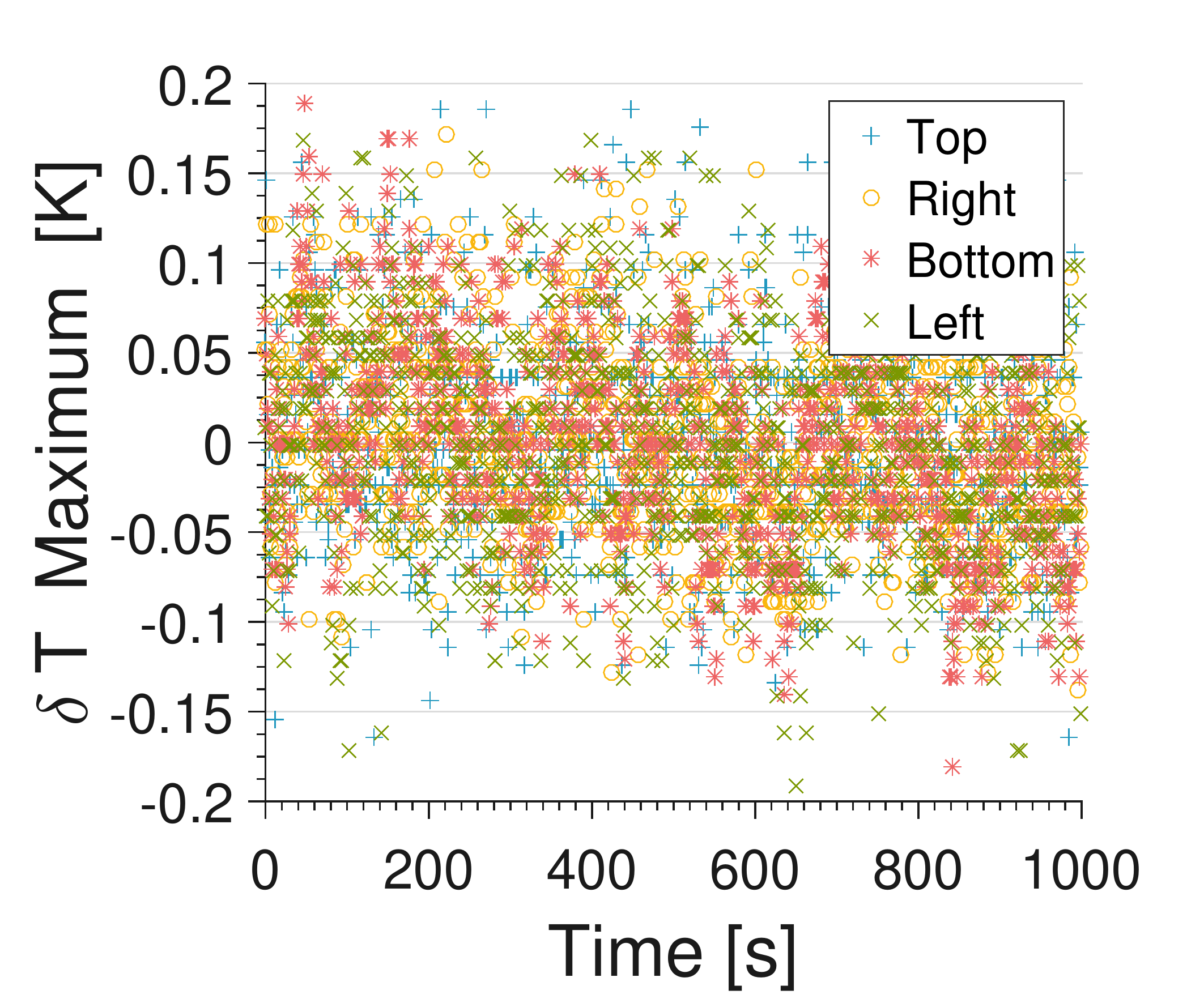}}{0.4cm}{1.4cm}     
        \end{subfigure}
  \caption{Fluctuation of the (a) mean, (b) minimum and (c) maximum temperature over time for each frame element.}\label{fig:temp_time}
\end{figure}

We define the temperature fluctuation $\delta T$ as the variation from the line profile which has been adjusted for its temporal average. We are able to control the average temperature control the average temperature fluctuation within $\left<\delta T \right>=$\SI{0.05}{\kelvin}.
The temperature fluctuation of the maximum, minimum and mean temperature over time is shown in figure \ref{fig:temp_time}. Additional asymmetry of the temperature curves stems from the surface contact of the heating elements as the surface pressure varies slightly.
The measured maximum heating temperature of the bottom segment is $T_{bot,max}=$\SI{55}{\celsius} and the minimum temperature of the top element is achievable $T_{top,min}=$\SI{15}{\celsius} with the bottom element at $T_{bot,max}$.

\section{Summary and Outlook}
A new apparatus for experiments on 2D convection has been developed, using thin liquid films. It consists of a fully transparent, sealable measurement chamber and a variable frame assembly. The principle application of the setup lies in the determination of the statistics of reversals in Rayleigh-B\'enard convection, where the aspect ratio $\Gamma$ can be effortlessly varied over a wide range. It thus allows the analysis of large scale convection reversals with varying geometry. The setup combines know-how from equilibrium state TLF experiments and classic Rayleigh-Bénard cells. The nonintrusive measurement technique CIV is applied to the system for measuring velocity profiles through the front observation window. The parameter space for turbulent TLF experiments has been extended due to the large aspect ratio variation and temperature range.

As mentioned above, reversals in thin liquid films have an increased turn over period compared to classic Rayleigh-Bénard setups. Therefore, it is possible to gather reliable statistics in a shorter amount of time or increase the accuracy with longer observation times. Results will be published elsewhere.

Another frontier in TLF convection experiments is the measurement of convection in the equilibrium phase (NBF or CBF), as no visible light is reflected from these films. Currently, the only way to get information on the velocity field is trough residual domains of thicker film which are advected and act as tracers.  Any PIV technique is even more out of the question at film thicknesses below \SI{50}{\nano\metre}. Two viable options exist to image motion in black films,
which will be explored: If the film is in the meta-stable Common black
film phase, the use of commercially available fluorescent particles
\cite{TSI2012Particles} may be possible. Otherwise fluorescent
molecules like manufactured liposomes are an option
\cite{singh2001fluorescent}.

We conclude that we have presented a dedicated device for the measurement of reversals in Rayleigh/B\'enard convection. Beyond this very specific use, general experiments for 2D liquids can be performed, we consider, however, the perspective for measurements on nonequilibrium thin-films as very promising, too. Controlled measurements in this direction are very complicated, and we have built a well-designed  setup where all relevant thermodynamic quantities can be controlled, including chemistry.


\begin{thebibliography}{10}

\bibitem{ahlers2009heat}
G.~{Ahlers}, S.~{Grossmann}, and D.~{Lohse}.
\newblock {Heat transfer and large scale dynamics in turbulent
  Rayleigh-B{\'e}nard convection}.
\newblock {\em Reviews of Modern Physics}, 81:503--537, April 2009.

\bibitem{LohseXia2010RBC}
D.~{Lohse} and K.-Q. {Xia}.
\newblock {Small-Scale Properties of Turbulent Rayleigh-B{\'e}nard Convection}.
\newblock {\em Annual Review of Fluid Mechanics}, 42:335--364, January 2010.

\bibitem{Sugiyama-Ni-Stevens-Chan-Zhou-Xi-Sun-Grossmann-Xia-Lohse-10}
Kazuyasu Sugiyama, Rui Ni, R.J~A.M. Stevens, T.~S.Chan, S.-Q. Zhou, H.-D. Xi,
  C.~Sun, S.~Grossmann, K.-Q. Xia, and D.~Lohse.
\newblock Flow reversals in thermally driven turbulence.
\newblock {\em Phys. Rev. Lett.}, 105:034503--1--4, 2010.

\bibitem{mishra2011dynamics}
Pankaj~Kumar Mishra, AK~De, MK~Verma, and V~Eswaran.
\newblock Dynamics of reorientations and reversals of large-scale flow in
  rayleigh--b{\'e}nard convection.
\newblock {\em Journal of Fluid Mechanics}, 668:480--499, 2011.

\bibitem{PhysRevLett.102.144503}
Fran{\c c}ois P\'etr\'elis, St\'ephan Fauve, Emmanuel Dormy, and Jean-Pierre
  Valet.
\newblock Simple mechanism for reversals of earth's magnetic field.
\newblock {\em Phys. Rev. Lett.}, 102(14):144503, 4 2009.

\bibitem{araujo2005wind}
Francisco~Fontenele Araujo, Siegfried Grossmann, and Detlef Lohse.
\newblock Wind reversals in turbulent rayleigh-b{\'e}nard convection.
\newblock {\em Physical review letters}, 95(8):084502, 2005.

\bibitem{martin1998spectra}
B.~K. Martin, X-L. Wu, W.~I. Goldburg, and M.~A. Rutgers.
\newblock {Spectra of decaying turbulence in a soap film}.
\newblock {\em Phys. Rev. Lett.}, 80(18):3964--3967, 1998.

\bibitem{zhang2005density}
J.~Zhang, XL~Wu, and K.Q. Xia.
\newblock {Density fluctuations in strongly stratified two-dimensional
  turbulence}.
\newblock {\em Physical review letters}, 94(17):174503, 2005.

\bibitem{seychelles2010intermittent}
F.~Seychelles, F.~Ingremeau, C.~Pradere, and H.~Kellay.
\newblock From intermittent to nonintermittent behavior in two dimensional
  thermal convection in a soap bubble.
\newblock {\em Physical review letters}, 105(26):264502, 2010.

\bibitem{exerowa1998foam}
D.~Exerowa and P.M. Kruglyakov.
\newblock {\em {Foam and foam films: theory, experiment, application}}.
\newblock Elsevier, New York, 1998.

\bibitem{ivanov1988thin}
IB~Ivanov.
\newblock {\em Thin liquid films: fundamentals and applications}, volume~29 of
  {\em surfactant science series}.
\newblock Marcel Dekker, Inc., 1988.

\bibitem{PhysRevE.76.056306}
Naveen Tiwari, Zoltan Mester, and Jeffrey~M. Davis.
\newblock Stability and transient dynamics of thin liquid films flowing over
  locally heated surfaces.
\newblock {\em Phys. Rev. E}, 76(5):056306, 11 2007.

\bibitem{stubenrauch2004stability}
C.~Stubenrauch and R.~Miller.
\newblock Stability of foam films and surface rheology: an oscillating bubble
  study at low frequencies.
\newblock {\em The Journal of Physical Chemistry B}, 108(20):6412--6421, 2004.

\bibitem{bergeron1999forces}
V.~Bergeron.
\newblock Forces and structure in thin liquid soap films.
\newblock {\em Journal of Physics: Condensed Matter}, 11:R215, 1999.

\bibitem{derjaguin1989theory}
B.V. Derjaguin.
\newblock {\em {Theory of stability of colloids and thin films}}.
\newblock Consultants Bureau New York and London, 1989.

\bibitem{reiter1998artistic}
G.~Reiter.
\newblock The artistic side of intermolecular forces.
\newblock {\em Science}, 282(5390):888, 1998.

\bibitem{kellay2011turbulence}
H.~Kellay.
\newblock Turbulence: Thick puddle made thin.
\newblock {\em Nature Physics}, 7(4):279--280, 2011.

\bibitem{yunker2011suppression}
P.J. Yunker, T.~Still, M.A. Lohr, and AG~Yodh.
\newblock Suppression of the coffee-ring effect by shape-dependent capillary
  interactions.
\newblock {\em Nature}, 476(7360):308--311, 2011.

\bibitem{davey2010enantiomer}
S.~Davey.
\newblock Enantiomer separation: Selective soap-films.
\newblock {\em Nature Chemistry}, 2010.

\bibitem{Prudhomme-Khan-96}
Eds. Prud'homme R.~K., Khan S.~A.
\newblock {\em Foams: Theory, Measurements, and Applications}.
\newblock Dekker, N.Y., 1996.

\bibitem{vermant2011fluid}
J.~Vermant.
\newblock Fluid mechanics: When shape matters.
\newblock {\em Nature}, 476(7360):286--287, 2011.

\bibitem{Verwey-Overbeek-48}
E.~J.~W. Verwey and J.~T.~G. Overbeek.
\newblock {\em {The Theory of the Stability of Lipophobic Colloids}}.
\newblock Elsevier, Amsterdam, 1948.

\bibitem{jones1966stability}
M.N. Jones, K.J. Mysels, and P.C. Scholten.
\newblock {Stability and some properties of the second black film}.
\newblock {\em Transactions of the Faraday Society}, 62:1336--1348, 1966.

\bibitem{israelachvili1991intermolecular}
J.N. Israelachvili.
\newblock {\em {Intermolecular and surface forces}}.
\newblock Academic press London, 1991.

\bibitem{Ivanov-80}
I.~B. Ivanov.
\newblock Foams: Theory, measurements, and applications.
\newblock {\em Pure Appl. Chem.}, 52:1241--1262, 1980.

\bibitem{zhang2005velocity}
J.~Zhang and XL~Wu.
\newblock {Velocity intermittency in a buoyancy subrange in a two-dimensional
  soap film convection experiment}.
\newblock {\em Physical review letters}, 94(23):234501, 2005.

\bibitem{seychelles2008thermal}
F.~Seychelles, Y.~Amarouchene, M.~Bessafi, and H.~Kellay.
\newblock {Thermal Convection and Emergence of Isolated Vortices in Soap
  Bubbles}.
\newblock {\em Phys. Rev. Lett.}, 100(14):144501, 2008.

\bibitem{breward2002drainage}
CJW Breward and PD~Howell.
\newblock {The drainage of a foam lamella}.
\newblock {\em Journal of Fluid Mechanics}, 458:379--406, 2002.

\bibitem{howell2005absence}
PD~Howell and HA~Stone.
\newblock On the absence of marginal pinching in thin free films.
\newblock {\em European Journal of Applied Mathematics}, 16(5):569--582, 2005.

\bibitem{muruganathan2004foam}
RM~Muruganathan, R~Krustev, H-J M{\"u}ller, H~M{\"o}hwald, B~Kolaric, and R~v
  Klitzing.
\newblock {Foam films stabilized by dodecyl maltoside. 1. Film thickness and
  free energy of film formation.}
\newblock {\em Langmuir: the ACS journal of surfaces and colloids},
  20(15):6352, 2004.

\bibitem{thess2001barrel}
A.~Thess, F.~Busse, R.~Du~Puits, C.~Resagk, and A.~Tilgner.
\newblock The barrel of ilmenau: a novel facility for experiments on high
  rayleigh number convection.
\newblock In {\em APS Division of Fluid Dynamics Meeting Abstracts}, volume~1,
  2001.

\bibitem{PhysRevE.75.016302}
Ronald du~Puits, Christian Resagk, and Andr\'e Thess.
\newblock Breakdown of wind in turbulent thermal convection.
\newblock {\em Phys. Rev. E}, 75(1):016302, 1 2007.

\bibitem{resagk2006oscillations}
C.~Resagk, R.~Du~Puits, A.~Thess, F.V. Dolzhansky, S.~Grossmann, F.F. Araujo,
  and D.~Lohse.
\newblock Oscillations of the large scale wind in turbulent thermal convection.
\newblock {\em Physics of Fluids}, 18:095105, 2006.

\bibitem{PhysRevLett.95.084502}
Francisco~Fontenele Araujo, Siegfried Grossmann, and Detlef Lohse.
\newblock Wind reversals in turbulent rayleigh-b\'enard convection.
\newblock {\em Phys. Rev. Lett.}, 95(8):084502, 8 2005.

\bibitem{bureau2010nonlinear}
L.~{Bureau}.
\newblock {Nonlinear Rheology of a Nanoconfined Simple Fluid}.
\newblock {\em Physical Review Letters}, 104(21):218302, May 2010.

\bibitem{stevens2009transitions}
R.~J.~A.~M. {Stevens}, J.-Q. {Zhong}, H.~J.~H. {Clercx}, G.~{Ahlers}, and
  D.~{Lohse}.
\newblock {Transitions between Turbulent States in Rotating Rayleigh-B{\'e}nard
  Convection}.
\newblock {\em Physical Review Letters}, 103(2):024503, July 2009.

\bibitem{PhysRevE.83.036307}
Takatoshi Yanagisawa, Yasuko Yamagishi, Yozo Hamano, Yuji Tasaka, and Yasushi
  Takeda.
\newblock Spontaneous flow reversals in rayleigh-b\'enard convection of a
  liquid metal.
\newblock {\em Phys. Rev. E}, 83(3):036307, 3 2011.

\bibitem{CambridgeJournals:7996276}
P.~K. MISHRA, A.~K. DE, M.~K. VERMA, and V.~ESWARAN.
\newblock Dynamics of reorientations and reversals of large-scale flow in
  rayleigh-b\'enard convection.
\newblock {\em Journal of Fluid Mechanics}, 668:480--499, 2011.

\bibitem{mysels1959soap}
K.J. Mysels, S.~Frankel, and K.~Shinoda.
\newblock {\em {Soap films: studies of their thinning and a bibliography}}.
\newblock Pergamon Press, 1959.

\bibitem{couder1989hydrodynamics}
Y.~Couder, J.M. Chomaz, and M.~Rabaud.
\newblock {On the hydrodynamics of soap films}.
\newblock {\em Physica D}, 37(1-3):384--405, 1989.

\bibitem{chomaz2001dynamics}
J.M. Chomaz.
\newblock {The dynamics of a viscous soap film with soluble surfactant}.
\newblock {\em Journal of Fluid Mechanics}, 442:387--409, 2001.

\bibitem{bruinsma1995theory}
R.~Bruinsma.
\newblock {Theory of hydrodynamic convection in soap films}.
\newblock {\em Physica A Statistical Mechanics and its Applications},
  216:59--76, 1995.

\bibitem{kellay1995experiments}
H.~Kellay, X-L Wu, and W.~I. Goldburg.
\newblock {Experiments with turbulent soap films}.
\newblock {\em Phys. Rev. Lett.}, 74(20):3975--3978, 1995.

\bibitem{Frisch-95}
U.~Frisch.
\newblock {\em Turbulence: The legacy of A. N. Kolmogorov}.
\newblock Cambridge Univ. Press, Cambridge, UK, 1995.

\bibitem{kellay2004dispersion}
H.~Kellay.
\newblock {Dispersion in the enstrophy cascade of two-dimensional decaying grid
  turbulence}.
\newblock {\em Physical Review E}, 69(3):36305, 2004.

\bibitem{jullien1999richardson}
M.C. Jullien, J.~Paret, and P.~Tabeling.
\newblock Richardson pair dispersion in two-dimensional turbulence.
\newblock {\em Physical review letters}, 82(14):2872--2875, 1999.

\bibitem{vorobieff1999soap}
P.~Vorobieff, M.~Rivera, and RE~Ecke.
\newblock {Soap film flows: Statistics of two-dimensional turbulence}.
\newblock {\em Physics of Fluids}, 11:2167, 1999.

\bibitem{rivera1998turbulence}
M.~Rivera, P.~Vorobieff, and R.E. Ecke.
\newblock {Turbulence in flowing soap films: Velocity, vorticity, and thickness
  fields}.
\newblock {\em Phys. Rev. Lett.}, 81(7):1417--1420, 1998.

\bibitem{shakeel2007decaying}
T.~Shakeel and P.~Vorobieff.
\newblock {Decaying turbulence in soap films: energy and enstrophy evolution}.
\newblock {\em Experiments in Fluids}, 43(1):125--133, 2007.

\bibitem{schnipper2009vortex}
T.~Schnipper, A.~Andersen, and T.~Bohr.
\newblock {Vortex wakes of a flapping foil}.
\newblock {\em Journal of Fluid Mechanics}, 633:411--423, 2009.

\bibitem{ChemSocRev39-2010}
Albert van~den Berg, Harold~G Craighead, and Peidong Yang.
\newblock From microfluidic applications to nanofluidic phenomena.
\newblock {\em Chem. Soc. Rev.}, 39(3):899--1220, 2010.

\bibitem{Physics.3.73}
Subhalakshmi~Kumar Steve~Granick, Sung Chul~Bae and Changqian Yu.
\newblock Confined liquid controversies near closure?
\newblock {\em Physics}, 3:73, 8 2010.

\bibitem{bocquet_nanofluidics_2010}
Lyderic Bocquet and Elisabeth Charlaix.
\newblock Nanofluidics, from bulk to interfaces.
\newblock {\em Chem. Soc. Rev.}, 39(3):1073--1095, 2010.

\bibitem{PhysRevLett.105.106101}
Shah~H. Khan, George Matei, Shivprasad Patil, and Peter~M. Hoffmann.
\newblock Dynamic solidification in nanoconfined water films.
\newblock {\em Phys. Rev. Lett.}, 105(10):106101, 8 2010.

\bibitem{isenberg1992science}
C.~Isenberg.
\newblock {\em {The science of soap films and soap bubbles}}.
\newblock Dover Pubns, 1992.

\bibitem{rutgers2001conducting}
MA~Rutgers, XL~Wu, and WB~Daniel.
\newblock {Conducting fluid dynamics experiments with vertically falling soap
  films}.
\newblock {\em Review of Scientific Instruments}, 72:3025, 2001.

\bibitem{zhang1996electrolyte}
L.~Zhang, P.~Somasundaran, and C.~Maltesh.
\newblock {Electrolyte Effects on the Surface Tension and Micellization of
  n-Dodecyl-$\beta$-d-Maltoside Solutions}.
\newblock {\em Langmuir}, 12(10):2371--2373, 1996.

\bibitem{winkler2013mixing}
M.~Winkler and M.~Abel.
\newblock Mixing in thermal convection of very thin free-standing films.
\newblock {\em Physica Scripta Volume T}, 155(1):014020, July 2013.

\bibitem{winkler2013exponentially}
M.~{Winkler}, G.~{Kofod}, R.~{Krastev}, S.~{St{\"o}ckle}, and M.~{Abel}.
\newblock Exponentially fast thinning of nanoscale films by turbulent mixing.
\newblock {\em Physical Review Letters}, 110(9):094501, March 2013.

\bibitem{cheung1996diffusion}
C.~Cheung, YH~Hwang, X.~Wu, and HJ~Choi.
\newblock Diffusion of particles in free-standing liquid films.
\newblock {\em Physical review letters}, 76(14):2531--2534, 1996.

\bibitem{vorobieff2001imaging}
P.~Vorobieff, M.~Rivera, and R.E. Ecke.
\newblock {Imaging 2D turbulence}.
\newblock {\em Journal of Visualization}, 3(4):323--330, 2001.

\bibitem{atkins2010investigating}
L.J. Atkins and R.C. Elliott.
\newblock Investigating thin film interference with a digital camera.
\newblock {\em American Journal of Physics}, 78:1248, 2010.

\bibitem{Joosten-88}
J.G.H. Joosten.
\newblock {\em Light Scattering from thin liquid films}, volume~29 of {\em
  surfactant science series}.
\newblock Marcel Dekker, Inc., 1988.

\bibitem{princen1965optical}
HM~Princen and SG~Mason.
\newblock Optical interference in curved soap films.
\newblock {\em Journal of colloid science}, 20(5):453--463, 1965.

\bibitem{gharib1999visualization}
M.~Gharib and M.~Beizaie.
\newblock Visualization of two-dimensional flows by a liquid (soap) film
  tunnel.
\newblock {\em Journal of visualization}, 2(2):119--126, 1999.

\bibitem{yang2001interpretation}
T.S. Yang, C.Y. Wen, and C.Y. Lin.
\newblock {Interpretation of color fringes in flowing soap films}.
\newblock {\em Experimental Thermal and Fluid Science}, 25(3-4):141--149, 2001.

\bibitem{winkler2011droplet}
M.~Winkler and M.~Abel.
\newblock Droplet coalescence in 2d thermal convection of a thin film.
\newblock {\em Journal of Physics Conference Series}, 333(1):012018, December
  2011.

\bibitem{minkina2009infrared}
W.~Minkina and S.~Dudzik.
\newblock {\em Infrared Thermography}.
\newblock Wiley Online Library, 2009.

\bibitem{zhang2006thermal}
J.~Zhang, XL~Wu, and N.~Rashidnia.
\newblock {Thermal radiation and thickness fluctuations in freely suspended
  liquid films}.
\newblock {\em Physics of Fluids}, 18:085110, 2006.

\bibitem{zhang2009persistent}
J.~Zhang and XL~Wu.
\newblock Persistent skewness of a strongly active scalar.
\newblock {\em Physical Review E}, 79(4):045301, 2009.

\bibitem{martin1998double}
B.~Martin and X-L. Wu.
\newblock {Double-diffusive convection in freely suspended soap films}.
\newblock {\em Phys. Rev. Lett.}, 80(9):1892--1895, 1998.

\bibitem{TSI2012Particles}
TSI.
\newblock Seed particles fluorescent particles 10 nm 540-560nm 10 ml.
\newblock website, , (\url{http://www.tsi.com/productView.aspx?id=24418}),
  2012.

\bibitem{singh2001fluorescent}
A.K. Singh, E.B. Cummings, and D.J. Throckmorton.
\newblock Fluorescent liposome flow markers for microscale particle-image
  velocimetry.
\newblock {\em Analytical chemistry}, 73(5):1057--1061, 2001.

\end{thebibliography}

\end{document}